\documentclass[12pt]{article}
\usepackage{amsmath} 
\usepackage{multirow} 
\usepackage{amsfonts}
\usepackage{amssymb}
\def\hybrid{\topmargin -20pt    \oddsidemargin 0pt
        \headheight 0pt \headsep 0pt
        \textwidth 6.25in       
        \textheight 9.5in       
        \marginparwidth .875in
        \parskip 5pt plus 1pt   \jot = 1.5ex}
\hybrid
\numberwithin{equation}{section}
\numberwithin{table}{section}

\newcommand{\beq}{\begin{equation}}
\newcommand{\eeq}{\end{equation}}
\newcommand{\bea}{\begin{eqnarray}}
\newcommand{\eea}{\end{eqnarray}}
\newcommand{\ba}{\begin{array}}
\newcommand{\ea}{\end{array}}
\newcommand{\bt}{\begin{tabular}}
\newcommand{\et}{\end{tabular}}
\newcommand{\bc}{\begin{center}}
\newcommand{\ec}{\end{center}}

\newcommand{\Ox}{\Omega}

\newcommand{\cL}{\mathcal{L}}
\newcommand{\cS}{\mathcal{S}}
\newcommand{\cK}{\mathcal{K}}
\newcommand{\cN}{\mathcal{N}}
\newcommand{\cW}{\mathcal{W}}
\newcommand{\cG}{\mathcal{G}}

\newcommand{\cF}{\mathcal{F}}

\newcommand{\KK}{\mathcal{K}}

\newcommand{\cM}{\mathcal M}

\newcommand{\IM}{\text{Im}\, \mathcal{M}}

\newcommand{\RM}{\text{Re}\, \mathcal{M}}
\newcommand{\I}{\text{Im}}
\newcommand{\R}{\text{Re}}

\newcommand{\Kcs}{K^{\text{cs}}}

\newcommand{\jb}{{\bar\jmath }}
\newcommand{\bj}{{\bar\jmath}}

\newcommand{\Kh}{{\hat{K}}}
\newcommand{\Lh}{{\hat{L}}}
\newcommand{\Ah}{{\hat{A}}}
\newcommand{\Bh}{{\hat{B}}}

\newcommand{\Mh}{{\hat{M}}}
\newcommand{\Nh}{{\hat{N}}}
\newcommand{\kh}{{\hat{k}}}
\newcommand{\ah}{{\hat{a}}}
\newcommand{\bh}{{\hat{b}}}
\newcommand{\ch}{{\hat{c}}}

\newcommand{\bbR}{\mathbb{R}}
\newcommand{\bbC}{\mathbb{C}}

\newcommand{\nn}{\nonumber}

\newcommand{\cref}{{\bf [check ref]}}

\newcommand{\Jc}{J_{\rm c}}
\newcommand{\Omegac}{\Omega_{\rm c}}
\newcommand{\cc}{c}
\newcommand{\CC}{C}

\begin{document}

\begin{titlepage}
\begin{center}


\vspace*{ 2cm}

{\large \bf  The effective action of Type IIA Calabi-Yau orientifolds}\footnote{%
Work supported by: DFG -- The German Science Foundation,
European RTN Program MRTN-CT-2004-503369  and the
DAAD -- the German Academic Exchange Service.}\\

\vskip 1.5cm

{\bf Thomas W.\ Grimm and Jan Louis}  \\
\vskip 0.3cm

{\em II. Institut f{\"u}r Theoretische Physik\\
Universit{\"a}t Hamburg\\
Luruper Chaussee 149\\
 D-22761 Hamburg, Germany}\\

{\tt  thomas.grimm@desy.de,   jan.louis@desy.de} \\

\end{center}

\vskip 2.5cm

\begin{center} {\bf ABSTRACT } \end{center}
The $N=1$ effective action for generic type IIA Calabi-Yau 
orientifolds in the presence of background fluxes is computed
from a Kaluza-Klein reduction. 
The K\"ahler potential, 
the gauge kinetic functions and the flux-induced superpotential 
are determined in terms of geometrical data of the
Calabi-Yau orientifold and the background fluxes. 
The moduli space is found to be a K\"ahler subspace of the
$N=2$ moduli space and shown to coincide 
with the moduli space arising in compactification
of M-theory on a specific class of $G_2$ manifolds.
The superpotential depends on all geometrical moduli and 
vanishes at leading order 
when background fluxes are turned off. 
The $N=1$ chiral coordinates linearize the appropriate
instanton actions such that instanton effects 
can lead to holomorphic corrections of the superpotential.
Mirror symmetry between type IIA and type IIB orientifolds
is shown to hold at the level of the effective action in
the large volume -- large complex structure limit.

\vfill

\noindent December 2004

\end{titlepage}

\section{Introduction}

Compactifications of type II string theories 
with space-time filling D-branes and
background fluxes are currently under investigation.
The reason is that they lead to phenomenologically interesting
string vacua both for particle physics as well as for cosmology 
\cite{reviewPP,reviewcosmo}.
If the string background includes a compact internal manifold $Y$,
consistency requires that apart from D-branes also negative tension 
objects have to be present.  Such objects are known as
orientifold planes and they arise when the string theory is modded out by a
discrete symmetry which includes parity reversal of the worldsheet
\cite{JP,JPbook,AD}. 

{} From a phenomenological point of view spontaneously broken
$N=1$ vacua are of particular interest. They can arise by first compactifying
type II string theories on specific orientifolds of Calabi-Yau
manifolds which preserve one of the two 
supersymmetries present in standard Calabi-Yau 
compactifications of type II theories  \cite{Ori,GKP,AAHV,BBKL,BH}. 
The D-branes can then be arranged
such that they preserve the same supersymmetry. 
These string vacua realize an  $N=1$ supersymmetry 
which is spontaneously  broken once background fluxes are turned on
\cite{Bachas,GVW,DSR}. 

In order to discuss the phenomenological properties of such vacua
in some detail it is essential to determine the low energy effective
action and in particular the couplings of the bulk moduli \cite{Soft}.
In this paper we focus on type IIA string theory compactified on
generic Calabi-Yau orientifolds 
and determine its  
low energy effective action
in terms of geometrical data of the 
Calabi-Yau orientifold and the background fluxes. 
Specifically we
determine the K\"ahler potential, the superpotential and the gauge-kinetic 
couplings by performing an appropriate Kaluza-Klein reduction.
We only discuss the couplings of the bulk moduli 
and leave their couplings to matter fields (and moduli) on the
D-branes for a future investigation.\footnote{We do include
D-branes for consistency but we freeze their moduli and matter fields
such that they do not appear in the low energy effective action.}

In standard $N=2$ Calabi-Yau compactifications the moduli space
consists of two components, a special K\"ahler manifold 
$\cM^{\rm K}$ and a quaternionic manifold $\cM^{\rm Q}$ 
\cite{Strominger,CdO,FS}.
The orientifold projection truncates the $N=2$ massless spectrum
and thus defines a K\"ahler submanifold in the 
moduli space of Calabi-Yau compactifications \cite{GKP,ADAF,BBHL,DFT,GL,Smet}.
This submanifold continues to be a product of two components
$\tilde\cM^{\rm K}\times\tilde\cM^{\rm Q}$.
For type IIA orientifolds we show that
$\tilde\cM^{\rm K}$ is again a special K\"ahler manifold 
characterized by a (truncated) holomorphic prepotential depending on the
complexified K\"ahler deformations of the Calabi-Yau orientifold.
The geometry of 
the K\"ahler submanifold $\tilde\cM^{\rm Q}$ inside $\cM^{\rm Q}$
turns out to be more involved.
This is due to the fact that the orientifold projection is
anti-holomorphic and destroys the complex structure
on the space of complex structure deformations.
Instead the complex structure of $\tilde\cM^{\rm Q}$
combines the type IIA RR~three-form $C_3$ with $\R \Omega$ to form
a `new' three-form $\Omegac = C_3 +2i \R(C\Omega)$ 
where $C$ is related to the inverse dilaton. 
The K\"ahler coordinates  of $\tilde\cM^{\rm Q}$
turn out to be half of the periods of $\Omega_c$
and the resulting geometry is similar 
to the geometry of the moduli space of Lagrangian submanifold
as discussed in ref.\ \cite{Hitchin2}. The K\"ahler potential
of $\tilde\cM^{\rm Q}$ encodes the dynamics of  $\R(C\Omega)$.

Once background fluxes are turned on a superpotential $W$ is generated.
Also $W$ decomposes into the sum of two terms analogously to the split
of the K\"ahler geometry. As a consequence $W$ depends on all
geometric moduli of the Calabi-Yau orientifold. For $N=2$ type IIA
compactification this has also been observed recently in ref.\ 
\cite{KachruK}.
The flux-induced superpotential receives further corrections from
worldsheet and $D$-brane instantons. We do not compute such corrections here
but observe that the chiral coordinates of $\tilde\cM^{\rm Q}$
are precisely such that they linearize the $D2$-brane instanton action
and hence holomorphic corrections to the superpotential are possible.

The type IIA orientifold compactifications considered in this paper
are closely related to M-theory compactifications on a
special class of $G_2$ manifolds \cite{HM,KMcG}. We show that 
the effective action of $G_2$ compactifications determined in
refs.\ \cite{PT,HM,Gukov,AS,Hitchin1,GPap,BW} indeed reduces 
in an appropriate limit to the effective action computed 
in this paper. This gives an alternative view on
the geometry of $\tilde\cM^{\rm Q}$ since it can also be understood
as a certain limit of the $G_2$ moduli space. In particular,
the definition of $\Omegac$ appear very naturally from an
M-theory perspective.

Type IIA and type IIB compactification on Calabi-Yau threefolds
are equivalent as a consequence of mirror symmetry \cite{HKT}.
In terms of the low energy effective action this implies that 
the two holomorphic $N=2$ prepotentials
of the special K\"ahler manifold 
$\cM^{\rm K}$ and the quaternionic manifold $\cM^{\rm Q}$ are equal.
In this way mirror symmetry computes the worldsheet instanton corrections
geometrically.
One expects mirror symmetry to be  also present 
in the orientifold versions of such compactifications \cite{AAHV,BH}. 
However, in this case the corrections are more difficult to control
since one 
can only rely on $N=1$ supersymmetry. Furthermore, since 
$\tilde \cM^{\rm Q}$ is not a special K\"ahler manifold
its geometry is no longer encoded in a holomorphic function and 
hence determining the corrections is less straightforward.
We take a purely supergravity point of view and compare 
the effective action computed in this paper with the type IIB
mirror actions determined in \cite{GL}.
Within this framework we find that for 
$\tilde\cM^{\rm K}$ mirror symmetry acts exactly
as in $N=2$ and equates the two orientifold truncated
holomorphic prepotentials. For
$\tilde \cM^{\rm Q}$ the situation is considerably more involved
and depending on the orientifold projection two inequivalent
mirrors do appear. 
On the  type IIA side
they correspond to two possible
sets of special coordinates while in type IIB they give rise to 
$O3/O7$ planes or $O5/O9$ planes.

The paper is organized as follows. To set the stage 
we briefly review the compactification of type IIA on a Calabi-Yau manifold 
in section \ref{revIIA}. 
In section \ref{IIAorientifolds} we turn to the discussion of type IIA
Calabi-Yau orientifolds. The orientifold projection is introduced in section \ref{spectrum} and the resulting  four-dimensional $N=1$ spectrum
is determined. In section \ref{eff_act}
we calculate the effective action by 
performing a Kaluza-Klein reduction while additionally imposing the orientifold constraints.
To bring the effective action in 
the standard $N=1$ form we determine the K\"ahler coordinates,
the K\"ahler potential and gauge-kinetic couplings in
section \ref{eff_supform}. 

In section \ref{fluxes_sec} we redo the reduction starting from massive type IIA supergravity \cite{Romans} while 
switching on the full set of possible NS and RR fluxes. 
This induces a superpotential for all geometric moduli which 
we determine explicitly. 
Furthermore we discuss contributions to the superpotential 
due to $D2$ instantons. 
By using the BPS conditions we show that the $D2$ instanton action 
becomes linear in the $N=1$ coordinates, which in fact 
is a generic feature of all supersymmetric D-instantons 
in type II Calabi-Yau orientifolds.

The embedding of IIA Calabi-Yau orientifolds into an M-theory compactification on 
a special class of $G_2$ manifolds is discussed in section \ref{G2}. 
We match explicitly the 
moduli spaces, gauge-couplings and parts of the flux superpotentials.

In section \ref{mirrorsec}, 
we discuss  
mirror symmetry for Calabi-Yau orientifolds and determine 
the necessary conditions
on the involutive symmetries of the mirror IIA and IIB orientifold theories. 
By specifying two types of 
special coordinates on the IIA side, we are able to identify the large complex 
structure limit of IIA orientifolds with the large volume limits of IIB orientifolds
with $O3/O7$ and $O5/O9$ planes. 

Section~\ref{Conclusion} contains our conclusions and some technical aspects
of our analysis are presented in four appendices.
Appendix~A briefly reviews the special geometry of the Calabi-Yau
moduli space. In appendix~B we summarize $N=1$ supergravities with
several linear multiplets  
as they are relevant in the computation of the effective action.
Appendix~C contains the details of the reduction of the quaternionic
manifold $\cM^{\rm Q}$ for an arbitrary symplectic basis of $H^3$.
Finally, appendix~D relates the geometry of $\tilde\cM^{\rm Q}$
to the moduli space of supersymmetric Lagrangian submanifolds
of Calabi-Yaus following \cite{Hitchin2}.

\section{Type IIA compactified on Calabi-Yau threefolds}
\label{revIIA}

In order to set the stage for the orientifold compactifications we 
briefly review  the compactification of type IIA supergravity
on a Calabi-Yau threefold $Y$ in this section \cite{BCF}. 
Since our main concern is the 
$N=2$ geometry of the moduli space we do not review the effective action where
in addition background fluxes are turned on  \cite{LM,KachruK}.

We start from the ten-dimensional type IIA supergravity action in the 
Einstein frame given by
\bea \label{10dact}
  S^{(10)}_{IIA} &=& \int -\tfrac{1}{2}\hat R*\mathbf{1} -\tfrac{1}{4} d\hat \phi\wedge * d\hat \phi
  -\tfrac{1}{4} e^{-\hat \phi}\hat H_3 \wedge *\hat H_3 
  -\tfrac{1}{2} e^{\frac{3}{2} \hat \phi}\hat F_2 \wedge *\hat F_2 \nn \\
  && -\tfrac{1}{2} e^{\frac{1}{2} \hat \phi}\hat F_4 \wedge *\hat F_4 
  + \cL_{\rm top}\ ,
\eea
where
\bea
  \cL_{\rm top}&=& -\tfrac{1}{2}\Big[ \hat B_2 \wedge d\hat C_3 \wedge d\hat C_3\  
                                   -(\hat B_2)^2 \wedge d\hat C_3
                                   \wedge d\hat A_1 \Big]\ ,
\eea
and the field strengths are defined as 
\bea \label{defHFF1}
  \hat H_3 = d \hat B_2\ , \quad \hat F_2 = d\hat A_1\ , \quad 
  \hat F_4 = d\hat C_3 - \hat A_1 \wedge \hat H_3\ .
\eea
The dilaton $\hat \phi$, the ten-dimensional metric $\hat g$ and the two-form $\hat B_2$ are the massless
fields in the NS sector, 
while the one- and three-forms $\hat A_1,\hat C_3$ arise in 
the RR sector.\footnote{We use a `hat'
to denote ten-dimensional quantities and omit it for 
four-dimensional fields.}

By compactifying this theory on a Calabi-Yau threefold $Y$ one obtains 
an $N=2$ theory in four space-time dimensions ($D=4$) where the zero
modes of $Y$ assemble into  massless  $N=2$ multiplets.
These zero modes are in one-to-one correspondence with harmonic forms
on $Y$ and thus their multiplicity is counted by the dimension
of the non-trivial cohomologies $H^{(1,1)}$ and $H^{(1,2)}$.
More precisely, one takes the ten-dimensional metric to be block diagonal
\beq
  ds^2\ = \ \eta_{\mu \nu}(x)\, dx^\mu dx^\nu + g_{i \jb}(y)\, dy^i dy^\jb\ ,
\eeq
where $\eta_{\mu \nu},\mu,\nu=0,\ldots,3$ 
is a four-dimensional Minkowski metric and 
$g_{i \jb},i,\jb=1 \ldots 3$ is a Calabi-Yau metric. 
Accordingly  we expand 
the ten-dimensional gauge-potentials introduced in \eqref{defHFF1} in 
terms of harmonic 
forms on $Y$
\bea \label{fieldexp}
  \hat A_1 &=& A^0(x)\ ,\qquad 
\hat B_2\, = \, B_2(x) +  b^A(x) \, \omega_A\ ,\quad  
       A\ =\ 1,\ldots, h^{(1,1)}\ ,\\
  \hat C_3 &=& 
A^A(x) \wedge \omega_A + \,
              \xi^\Kh(x)\, \alpha_\Kh - \tilde \xi_\Kh(x)\, \beta^\Kh\ , \quad \Kh\ =\ 0,\ldots, h^{(2,1)}\ . \nn 
\eea  
Here $b^A,\xi^\Kh,\tilde \xi_\Kh$ are four-dimensional scalars, 
$A^0,A^A$ are one-forms and  $B_2$ is a two-form. 
The harmonic forms $\omega_A$ form a  basis of
$H^{(1,1)}(Y)$ on the internal Calabi-Yau $Y$
while  the
$(\alpha_\Kh,\beta^\Kh)$ form a real symplectic basis of $H^3(Y)$ in
that
they satisfy 
\beq\label{symplectic}
  \int \alpha_\Kh \wedge \beta^\Lh =\delta_\Kh^\Lh\ ,
\eeq 
with all other intersections vanishing. 
The ten-dimensional one-form $\hat A_1$
only contains a four-dimensional one-form $A^0$ in the expansion
\eqref{fieldexp} since a Calabi-Yau 
threefold has no harmonic one-forms. 

The four-dimensional massless modes are completed by also 
taking deformations 
of the Calabi-Yau metric $g_{i\jb}$ into account.
These deformations are divided into the deformations of the K\"ahler
form $J$ and deformations of the complex structure.
The former correspond to $h^{(1,1)}$ real  scalars $v^A$ while the later
are $h^{(1,2)}$ complex scalars $z^K, K = 1,\ldots,h^{(2,1)}$.\footnote{%
Note that $\hat K$ introduced in  
\eqref{fieldexp} takes one more value than $K$ in that it includes the zero.} 
Together with the fields defined in the expansion \eqref{fieldexp}
they assemble into a gravity multiplet $(g_{\mu\nu},A^0)$,
$h^{(1,1)}$ vector multiplets $(A^A, v^A, b^A)$,
$h^{(2,1)}$ 
hypermultiplets $(z^K,\xi^K,\tilde \xi_K)$ and one tensor multiplet 
$(B_2,\phi,\xi^0,\tilde \xi_0)$ where we only give the bosonic 
components.
Dualizing the two-form $B_2$ to a scalar $a$ results in one
further hypermultiplet. We summarize the bosonic spectrum in 
table~\ref{tab-compIIAspec}.

\begin{table}[h]
\begin{center}
\begin{tabular}{| c | c | c |} \hline
   \rule[-0.3cm]{0cm}{0.8cm} gravity multiplet  &
   $1$ & {\small $(g_{\mu \nu},A^0)$} 
   \\ \hline
   \rule[-0.3cm]{0cm}{0.8cm} vector multiplets &
   $h^{(1,1)}$ & {\small $(A^{A}, v^A,b^A)$}\\ \hline
   \rule[-0.3cm]{0cm}{0.8cm} hypermultiplets  &
   $h^{(2,1)}$ & 
   {\small $(z^K,\xi^K,\tilde \xi_K)$}\\ \hline
\rule[-0.3cm]{0cm}{0.8cm} tensor multiplet  &
   1 &
   {\small $(B_2,\phi,\xi^0,\tilde \xi_0)$} \\ \hline

\end{tabular}
\caption{\small \label{tab-compIIAspec}
\textit{ $N=2$ multiplets for Type IIA supergravity compactified on a Calabi-Yau manifold.}}
\end{center}
\end{table}
In order to display the low energy  effective action in the standard 
$N=2$ form one needs to redefine the field variables slightly.
One combines the real scalars $v^A, b^A$ into  complex fields 
$t^A$  and defines a four-dimensional dilaton $D$ according to\footnote{%
The fields $v^A$ are defined as the expansion coefficients of
the K\"ahler form $J$ in the string-frame $J=v^A\omega_A$
which is related to the K\"ahler form $J_E$ in the Einstein-frame
via $J = e^{\phi/2}J_E$.}
\beq \label{4d-dilaton}
t^A = b^A + i\, v^A\ , \qquad 
   e^{D} = e^{\phi} (\cK/6)^{-\frac{1}{2}}\ ,
\eeq
where $\cK= \int J \wedge J \wedge J = 6\, {\rm vol}(Y)$ 
is proportional to volume of $Y$ in the string-frame.
Inserting the field expansions \eqref{fieldexp} into \eqref{defHFF1}, \eqref{10dact}
and reducing the Riemann scalar $R$ by including the complex and K\"ahler 
deformations one ends up with the four-dimensional $N=2$ effective action  
\cite{N=2review,BCF,FS}
\bea \label{IIA-4}
  S^{(4)}_{\rm IIA} & = &\int -\tfrac12 R * \mathbf{1} 
                     +  \tfrac12 \I \cN_{\Ah \Bh}\, F^{\Ah} \wedge *
                     F^{\Bh}
                   +  \tfrac12 \R \cN_{\Ah \Bh}\, F^{\Ah} \wedge  F^{\Bh} \\
                    & & - G_{A B}\, dt^A \wedge *  d\bar t^B 
                        - h_{uv}\, d\tilde q^u \wedge * d\tilde q^v \ , \nn
\eea
where $F^{\Ah} = dA^{\Ah}$.

Let us first discuss 
the couplings of the hypermultiplet sector which are encoded in the
quaternionic metric $h_{uv}$. From the 
Kaluza-Klein reduction one obtains \cite{FS}
\bea \label{q-metr}
  h_{uv}\, d\tilde q^u\,  d\tilde q^v &=&  (dD)^2 + G_{K \bar L}\, dz^K d\bar z^L
                               +\tfrac{1}{4}e^{4D}\big(da -(\tilde \xi_\Kh d\xi^\Kh - \xi^\Kh d\tilde \xi_\Kh) \big)^2 \\
                              && - \tfrac{1}{2} e^{2D} (\text{Im}\; \cM)^{-1\ \Kh \Lh}
                                   \big(d\tilde \xi_\Kh - \cM_{\Kh \Nh} d\xi^\Nh \big)
                                   \big(d\tilde \xi_\Lh - \bar
                                   \cM_{\Lh \Mh} d\xi^\Mh \big)\ .\nn 
\eea
$G_{K \bar L}$ is the metric on the submanifold $\cM^{\rm cs}$
spanned by the complex structure
deformations $z^K$ and given by \cite{Tian,CdO}
\beq \label{chi_barchi}
  G_{K \bar L} = -\frac{\int_Y \chi_K \wedge \bar \chi_{ L}}{\int_Y \Ox \wedge \bar \Ox}\ .
\eeq
$\chi_K$ is a basis of $(2,1)$-forms related to the
variation of the three-form
$\Omega$ via Kodaira's formula
\beq \label{Kod-form}
\chi_K(z) = \partial_{z^K} \Omega(z)+ \Omega(z)\, \partial_{z^K}\Kcs   \ .
\eeq
With the help of \eqref{Kod-form} one shows that $G_{K \bar L}$ 
 is  a special K\"ahler metric determined by
the periods of $\Omega$ 
\beq\label{csmetric}
G_{K \bar L} = 
{\partial}_{z^K}\partial_{\bar z^L}
\  \Kcs\ , \qquad
 \Kcs = -\ln\Big[ i \int_Y \Ox \wedge \bar \Ox\Big] 
= -\ln i\Big[\bar Z^\Kh\mathcal{F}_\Kh - Z^\Kh\bar{\mathcal{F}}_\Kh \Big]
\ ,
\eeq
where the holomorphic periods  $Z^\Kh, \mathcal{F}_\Kh$ are defined as
\beq \label{pre-z}
Z^\Kh(z) = \int_Y \Omega(z) \wedge \beta^\Kh\ , \qquad 
\cF_\Kh(z) = \int_Y \Omega(z) \wedge \alpha_\Kh\ , 
\eeq
or in other words $\Omega$ enjoys the expansion
\beq\label{Omegaexp}
  \Omega(z) = Z^\Kh(z)\, \alpha_\Kh - \cF_\Kh(z)\, \beta^\Kh\ .
\eeq
$\cF_\Kh$ is the first derivative with respect to $Z^\Kh$ of 
a prepotential 
$\cF = \frac{1}{2} Z^\Kh \cF_\Kh$. 
(We briefly summarize the special geometry of the Calabi-Yau moduli space
in appendix~\ref{specialGeom}.)

$\Omega$ is only defined up to complex rescalings by a holomorphic function
$e^{-h(z)}$ which via \eqref{csmetric} 
also changes the K\"ahler potential by a K\"ahler transformation 
\beq\label{crescale}
\Omega\to\Omega\, e^{-h(z)}\ , \qquad \Kcs\to\Kcs + h +\bar h\ .
\eeq
This symmetry renders one of the periods (conventionally
denoted by $Z^0$) unphysical
in that one can always choose to fix a K\"ahler gauge and set $Z^0 = 1$. 
The complex structure
deformations can thus be identified with the remaining 
$h^{(1,2)}$ periods $Z^K$ by defining the special coordinates
$z^K = \frac{Z^K}{Z^0}$.

The complex coupling matrix $\cM_{\Kh \Lh}$ appearing in
\eqref{q-metr} 
depends on the complex structure deformations $z^K$ and is defined as
\cite{CDAF}
\bea \label{defM}
  \int \alpha_\Kh \wedge * \alpha_\Lh&=&-(\text{Im}\; \cM +(\text{Re}\; \cM)
  (\text{Im}\; \cM)^{-1}(\text{Re}\; \cM))_{\Kh \Lh}\ , \nn\\
  \int \beta^\Kh \wedge * \beta^\Lh &=&-(\text{Im}\; \cM)^{-1\ \Kh \Lh}\ ,  \\
  \int \alpha_\Kh\wedge * \beta^\Lh &=& 
  -((\text{Re}\; \cM)(\text{Im}\; \cM)^{-1})_{\Kh}^\Lh\ .  \nn
\eea
It can be calculated from the periods \eqref{pre-z} by using equation \eqref{gauge-c}.
Thus in the hypermultiplet sector all couplings are determined
by a holomorphic prepotential and such metrics have been called dual or special
quaternionic
\cite{CFGi,FS}.

Now let us turn to the couplings of the vector multiplets in the action
\eqref{IIA-4}.
The metric $G_{A \bar B}$ only depends on the $t^A$ (or rather
their imaginary parts) and is defined as \cite{Strominger,CdO}
\bea \label{Kmetric} 
  G_{A B} = \frac{3}{2\KK}
  \int_{Y}\omega_A \wedge *\omega_B = -\frac{3}{2}\left( \frac{\KK_{AB}}{\KK}-
  \frac{3}{2}\frac{\KK_A \KK_B}{\KK^2} \right)
= -\partial_{t^a} \partial_{\bar t^B} \ln \tfrac43\KK
\ .
\eea
We abbreviated the intersection numbers as follows
\begin{align}\label{int-numbers}
  \KK_{ABC} &= \int_{Y}\omega_A \wedge \omega_B \wedge \omega_C\ ,& 
  \KK_{AB}  &= \int_{Y}\omega_A \wedge \omega_B \wedge J 
= \KK_{ABC}v^C\ ,&  \\
  \KK_{A}   &= \int_{Y}\omega_A \wedge J \wedge J
=\KK_{ABC}v^Bv^C \ ,
& 
  \KK &= \int_{Y}J \wedge J \wedge J
 =\KK_{ABC}v^Av^Bv^C \ ,\nonumber
\end{align}
with $J = v^A \omega_A $ being the K\"ahler form of $Y$ in the string-frame. 
The metric \eqref{Kmetric} is again a special K\"ahler metric
in that the K\"ahler potential 
$K^{\rm K} = -\ln \frac43\KK$ is also determined 
by a prepotential $f(t)$ given in \eqref{pre-K} via \eqref{Kinz}. 

Finally, 
the gauge-kinetic coupling matrix $\cN_{\Ah \Bh}$ also depends on 
the scalars $t^A$ and is given explicitly in \eqref{def-cN}.
It can be calculated from a holomorphic prepotential as 
explained in appendix \ref{specialGeom}.

As we have just reviewed the $N=2$ moduli space 
has the local product structure 
\beq \label{N=2modsp}
  \cM^{\rm K} \times \cM^{\rm Q}\ ,
\eeq
where $\cM^{\rm K} $ is the special K\"ahler manifold spanned
by the scalars in the vector multiplets or in other words
the (complexified) deformations of the Calabi-Yau K\"ahler form
and $\cM^{\rm Q}$ is a dual quaternionic manifold spanned by
the scalars in the hypermultiplets.
$\cM^{\rm Q}$ has a special K\"ahler submanifold spanned by the 
complex structure deformations or in other words the geometric
Calabi-Yau moduli space has the structure
\beq \label{geom-mod}
  \cM^{\rm K} \times \cM^{\rm cs}\ ,
\eeq
where both factors are special K\"ahler manifolds of complex dimension $h^{2,1}$ and $h^{1,1}$
respectively.

This ends our short review of Calabi-Yau compactifications of type IIA
supergravity. Next we turn to its orientifold version which breaks
$N=2$ to $N=1$ and as a consequence truncates the massless spectrum.
This defines a K\"ahler submanifold inside the $N=2$ moduli space
\eqref{N=2modsp}. After determining the $N=1$ spectrum 
we are going to find this K\"ahler subspace.

\section{IIA orientifolds}
\label{IIAorientifolds}

After this brief review let us now turn to the main point of our paper
and compactify type
IIA supergravity on Calabi-Yau orientifolds. 
We first discuss the orientifold projection and the resulting
$N=1$ spectrum in section 3.1. In 3.2 we derive the effective action
from a Kaluza-Klein reduction or equivalently by truncating the
$N=2$ action of the previous section.
In 3.3 we find the appropriate chiral field variables which puts
the action into the standard $N=1$ form and
determine the K\"ahler potential and the gauge kinetic
function. In section 3.4 we redo the Kaluza-Klein reduction
using as the starting point the massive ten-dimensional IIA supergravity
of ref.\ \cite{Romans}.
We turn on background fluxes 
and determine the flux-induced superpotential.
We also include a brief discussion of 
possible instanton corrections to the superpotential.
Specifically we show that the $D2$ instanton action
becomes linear in the chiral $N=1$ coordinates 
and therefore holomorphic corrections to the superpotential
can be induced.

%
%

\subsection{The orientifold projection and the $N=1$ spectrum 
\label{spectrum}}

A Calabi-Yau orientifold is constructed from a Calabi-Yau manifold
by modding out a discrete symmetry  $\mathcal{O}$ which includes 
the world-sheet parity $\Omega_p$ combined with 
 the space-time fermion number in the left-moving sector
$(-1)^{F_L}$. In addition $\mathcal{O}$ can act non-trivially
on the Calabi-Yau manifold so that one has altogether 
\beq \label{oproj}
  \mathcal{O} = \Omega_p (-1)^{F_L} \sigma\ ,
\eeq
where $\sigma$ is an involutive symmetry 
of $Y$ (i.e.\ $\sigma^2=1$), acting trivially on 
the four flat dimensions.
If one insists on preserving $N=1$
supersymmetry $\sigma$ has to be anti-holomorphic and
isometric such that the K\"ahler form transforms as \cite{AAHV,BBKL,BH}
\beq \label{constrJ}
  \sigma^* J\ =\ -J\ , \quad 
\eeq
where $\sigma^*$ denotes the pullback of the map $\sigma$.
Compatibility of $\sigma$ with the Calabi-Yau condition 
$\Omega \wedge \bar \Omega \propto J \wedge J \wedge J$ 
implies that $\sigma$ also acts non-trivially  on the three-form $\Omega$ as
\beq \label{constrO}
  \sigma^* \Omega\ =\ e^{2i\theta} \bar \Omega \ , 
\eeq
where $e^{2i\theta}$ is a constant phase and we included a factor 2 for later convenience.

Type IIA orientifolds with anti-holomorphic involution generically admit $O6$ planes. This 
is due to the fact, that the fixed point set of $\sigma$ in $Y$ are  three-cycles 
$\Lambda_n$ supporting the internal part of the orientifold planes. 
These cycles are special Lagrangian submanifolds of $Y$ as an 
immediate consequences of \eqref{constrJ} and \eqref{constrO}
which implies \cite{HitchinLec}
\beq \label{OLagr}
  J|_{\Lambda_n} = 0\ , \qquad  \I(e^{-i\theta}\Omega)|_{\Lambda_n} = 0\ .
\eeq
In other words, they are calibrated with respect to 
$\R(e^{-i\theta}\Omega)$
\beq \label{calibr-O6}
\rm{vol}(\Lambda_n)\sim \int_{\Lambda_n}\R(e^{-i\theta}\Omega)\ ,
\eeq
where the overall normalization of $\Omega$ will be determined
in \eqref{Omeganorm}.\footnote{%
As we discuss in section \ref{fluxes_sec} this calibration condition plays a
central role when including corrections due to BPS $D2$ instantons.}
%


In order to determine the  $\mathcal{O}$-invariant states let us recall
that the ten-dimensional RR forms $\hat A_1$ and $\hat C_3$ 
are odd 
under $(-1)^{F_L}$ while all other fields are even. 
Under the worldsheet 
parity $\Omega_p$ on the other hand $\hat B_2, \hat C_3$ are odd
with all other fields being even.
As a consequence the 
$\mathcal{O}$-invariant states have to satisfy
\cite{BH}
\begin{equation} \label{fieldtransf}
\begin{array}{lcl}
\sigma^*  \hat \phi &=& \  \hat \phi\ , \\
\sigma^*   \hat g &=& \ \hat g\ , \\
\sigma^*   \hat B_2 &=& -  \hat B_2\ ,
\end{array}
\hspace{2cm}
\begin{array}{lcl}
\sigma^*   \hat A_1 &=&  -  \hat A_1\ , \\
\sigma^*   \hat C_3 &=& \  \hat C_3\ ,
\end{array}
\end{equation}
while the deformations of the Calabi-Yau metric are constrained
by \eqref{constrJ} and \eqref{constrO}.\footnote{%
Following the argument presented in 
\cite{BH} we note that the involution does not change
 under deformations of $Y$. 
This is due to its involutive property 
and the fact that we identify involutions which differ
by diffeomorphisms.
Therefore we fix an involution and restrict the deformation space by demanding 
\eqref{constrJ} and \eqref{constrO}. }

As we recalled in the previous section the massless modes are in one-to-one
correspondence with the harmonic forms on $Y$. The space of harmonic forms
splits under the involution $\sigma$ into even and odd eigenspaces
\beq \label{cohom-split}
   H^p(Y)\ =\ H^p_+ \oplus H^p_-\ \ .
\eeq
Depending on the transformation properties given in \eqref{fieldtransf}
the $\mathcal{O}$-invariant states reside either in $H^p_+$ or in $H^p_-$
and as a consequence the number of states is reduced.
We summarize all non-trivial cohomology groups including their basis elements 
in table \ref{CYObasis}.\\

\begin{table}[h]
\begin{center}
\begin{tabular}{| c || c | c| c | c | c | c |} \hline
   \rule[-0.3cm]{0cm}{0.9cm} cohomology group &  $\ H^{(1,1)}_+\ $ & 
   $\ H^{(1,1)}_-\ $ & $\ H^{(2,2)}_+\ $ & $\ H^{(2,2)}_-\ $ & $\ H^{(3)}_+\ $ & $\ H^{(3)}_-\ $
   \\ \hline
   \rule[-0.3cm]{0cm}{0.8cm} dimension &  $h^{(1,1)}_+$  & $h^{(1,1)}_- $  
                                       &  $h^{(1,1)}_-$  & $h^{(1,1)}_+$ 
                                       &  $h^{(2,1)}+1$  &  $h^{(2,1)}+1$ 
   \\ \hline
   \rule[-0.3cm]{0cm}{0.8cm} basis     & $\omega_\alpha$ & $\omega_a$
                                       & $\tilde \omega^a$ & $\tilde \omega^\alpha$
   & $ a_{\Kh}$ & $b^{\Kh}$ \\ \hline
\end{tabular}
\caption{\small \label{CYObasis}
\textit{Cohomology groups and their basis elements.}}
\end{center}
\end{table}

$\omega_\alpha, \omega_a$ denote
 even and odd $(1,1)$-forms while 
$\tilde\omega^\alpha, \tilde\omega^a$ denote odd and even
 $(2,2)$-forms. The number of even $(1,1)$-forms is equal to the number
of odd $(2,2)$-forms and vice versa since the 
volume form which is 
proportional to $J \wedge J \wedge J$ is odd and thus Hodge duality
demands $h^{(1,1)}_+ = h^{(2,2)}_- ,\ h^{(1,1)}_- = h^{(2,2)}_+$.
This can also be seen from the fact that the non-trivial
intersection numbers are 
\beq \label{basis-int}
  \int \omega_\alpha \wedge \tilde \omega^\beta =
  \delta^{\beta}_\alpha\ ,\quad \alpha,\beta = 1, \ldots, h^{(1,1)}_+\ ,
  \qquad 
  \int \omega_a \wedge \tilde \omega^b = \delta^{b}_a\ , \quad
  a,b=1,\ldots,h^{(1,1)}_-\ ,
\eeq
with all other pairings vanishing.
{}From the volume-form being odd 
one further infers $h^{(3,3)}_+=0,$ $h^{(3,3)}_-=1$ and 
$h^{(0,0)}_+=1,\ h^{(0,0)}_-=0$.

$H^{3}$ can be decomposed 
independently of the complex structure as
$H^{3}=H^3_+ \oplus H^3_-$ where 
the (real) dimensions of  both $H^3_+$ and $H^3_-$
is equal and given by $h^{3}_+ =h^{3}_-= h^{(2,1)}+1$.
Again this is a consequence of Hodge duality
together with the fact that the volume-form is odd.
It implies that for each element $a_\Kh \in H^3_+$ 
there is a dual element $b^\Lh \in H^3_-$
with the intersections
\beq \label{basis_ab}
 \int a_\Kh \wedge b^\Lh = \delta^\Lh_\Kh \ , \qquad 
 \Kh, \Lh = 0,\ldots, h^{(2,1)}\ .
\eeq
Compared to \eqref{symplectic} this amounts to a symplectic rotation
such that all $\alpha$-elements are chosen to be even and 
all $\beta$-elements are chosen to be odd but with the intersection
numbers unchanged.
The orientifold projection breaks this symplectic invariance 
or in other words fixes a particular symplectic gauge 
which groups all basis elements into even and odd. 
This in turn implies that the basis $(a_\Kh,b^\Kh)$ is only one possible choice. 
However, since the calculation simplifies considerably for this basis, we first restrict 
to this special case and later give the general results with calculations summarized in appendix~\ref{Geom-of-M}.

In the remainder of this subsection we determine the $N=1$ spectrum 
which survives the orientifold projections.
Let us first discuss the K\"ahler moduli. 
From the eqs. \eqref{constrJ} and \eqref{fieldtransf} we see that both
$J$ and $\hat B_2$ are odd and hence have to be expanded
in a basis $\omega_a$ of odd harmonic $(1,1)$-forms 
\beq \label{expJB}
  J\ =\ v^a(x)\, \omega_a\ ,\qquad  \hat B_2\ =\ b^a(x)\, \omega_a\ , \qquad a = 1,\ldots, h^{(1,1)}_-\ .
\eeq
In contrast to \eqref{fieldexp} the four-dimensional 
two-form $B_2$ gets projected out due to \eqref{fieldtransf} and the fact
that $\sigma$ acts trivially on the flat dimensions. 
$v^a$ and $b^a$ are space-time scalars and 
as in $N=2$ they can be combined into complex coordinates
\beq \label{def-t}
  t^a = b^a + i\,  v^a\ , \qquad \Jc = B_2 + i J\ ,
\eeq
where we have also introduced the complexified 
K\"ahler form $\Jc$.
We see that in terms of the field variables the same complex
structure is chosen as in $N=2$ but the dimension of the K\"ahler moduli
space is truncated from $h^{(1,1)}$ to $h^{(1,1)}_-$.

The number of complex structure deformations is similarly reduced since
\eqref{constrO} constrains the possible deformations.
To see this one performs a symplectic rotation on 
\eqref{Omegaexp} and expands $\Omega$ in the basis of
$H^p_+ \oplus H^p_-$, i.e.\ as\footnote{Let us stress that at this
point all $N=2$ relations are still intact since \eqref{Omegapm}
is just a specific choice of the standard $N=2$ basis \eqref{Omegaexp}.} 
\beq\label{Omegapm}
\Omega(z) = Z^\Kh(z)\, a_\Kh - \cF_\Lh(z)\, b^\Lh\ .
\eeq 
Inserted into \eqref{constrO} one finds 
\bea \label{Z=0}
   \I(e^{-i\theta} Z^\Kh)\ =\  0\ , \qquad
   \R(e^{-i\theta} \cF_\Kh )\ = 0\ .
\eea  
The first set of equations  are $h^{(2,1)}+1$ real conditions
for $h^{(2,1)}$ complex scalars $z^K$.
One of these equations is redundant due to the 
scale invariance \eqref{crescale} of $\Omega$.
More precisely, the phase of $e^{-h}$ can be used to 
trivially satisfy $\I(e^{-i\theta} Z^\Kh)= 0$ for one of the $Z^\Kh$.
Thus $\I(e^{-i\theta} Z^\Kh)=0$
projects out $h^{(2,1)}$ real scalars, i.e.\
half of the complex structure deformations. 
Furthermore, in section \ref{eff_act} we will see 
the remaining real complex structure deformations 
span a Lagrangian submanifold $\cM^{\rm cs}_\bbR$
with respect to the K\"ahler form
inside $\cM^{\rm cs}$. 
Note that the second set of equations in \eqref{Z=0}
$\R (e^{-i\theta} \cF_\Kh ) = 0$ 
should not be read as equations determining
the $z^K$ but is a constraint on the periods (or equivalently
the Yukawa couplings) of the Calabi-Yau
which has to be fulfilled in order to admit an involutive symmetry 
with the property \eqref{constrO}.\footnote{This can also be seen 
as conditions arising in consistent truncations of 
$N=2$ to $N=1$ theories as discussed in ref.\ \cite{ADAF}.}

As we have just discussed 
the complex rescaling \eqref{crescale}
is reduced to the freedom of a real rescaling by \eqref{constrO}. 
Under these transformations $\Omega$ and the K\"ahler potential $\Kcs$ 
change as
\beq \label{real_K}
  \Omega\to\Omega\, e^{-\R(h)}\ , \qquad \Kcs\to\Kcs + 2 \R(h)\ ,
\eeq
when restricted to $\cM^{\rm cs}_\bbR$. This freedom can be used to set one of 
the $\R (e^{-i\theta}Z^{\Kh})$ equal to one and tells us
that $\Omega$ depends only on $h^{(2,1)}$ real deformation parameters. 
However, it will turn out to be more
convenient to leave this gauge freedom intact and define 
a complex `compensator' $C=re^{-i\theta}$ with the transformation property
$C\to C e^{\R (h)}$.\footnote{This is reminiscent of the situation
encountered in the computation of the entropy of $N=2$ black holes
where it is also convenient to leave this scale invariance intact \cite{OSV}.}
Later on we will relate $r$ to 
the inverse of the four-dimensional dilaton 
so that the scale invariant function $C\Omega$ depends on 
$h^{(2,1)}+1$ real parameters. 
Using \eqref{Omegapm} $C\Omega$ enjoys
the expansion 
\beq \label{decompO}
  C \Omega\ =\ \R (C Z^\Kh)\, a_\Kh - i\I (C \cF_\Lh)\,  b^\Lh\ .
\eeq

We are left with the expansion of the ten-dimensional fields
$\hat A_1$ and $\hat C_3$ into  harmonic forms. 
{}From \eqref{fieldtransf} we learn that $\hat A_1$ is odd
and so together with the fact that
$Y$ posses no harmonic one-forms 
and $\sigma$ acts trivially on the flat dimensions
the entire $\hat A_1$ is projected out. This 
corresponds to the fact that the $N=2$ graviphoton $A^0$ is removed
from the gravity multiplet, 
which in $N=1$ only consists of the metric $g_{\mu \nu}$ as 
bosonic component.
Finally, $\hat C_3$ is even and thus can be expanded according to
\beq \label{form-exp}
  \hat C_3 = \cc_3(x) 
+ A^\alpha(x) \wedge \omega_\alpha + \CC_3\ ,\qquad
\CC_3 \equiv \xi^\Kh(x)\,  a_\Kh \ ,
\eeq
where $\xi^\Kh$ are $h^{(2,1)}+1$ real 
scalars, $A^\alpha$ are $h^{(1,1)}_+$ one-forms
and $\cc_3$ is a three-form in four dimensions.
$\cc_3$ contains no physical degree of freedom but as we will see 
in section~\ref{fluxes_sec} corresponds to a 
constant flux parameter in the superpotential.
The real scalars 
$\xi^\Kh$ have to combine with the 
$h^{(2,1)}$ real  complex structure deformations 
and the dilaton to form chiral multiplets.
In the next section we will find that the appropriate complex fields 
arise from the combination
\beq\label{Omegacdef}
  \Omegac\ =\ \CC_3 + 2i\R(C\Omega) \ .
\eeq
Expanding $\Omegac$  in a basis \eqref{basis_ab} of $H^3_+(Y)$ 
and using \eqref{decompO} and \eqref{form-exp} we have 
\beq\label{newO}
\Omegac\ =\ 2 N^\Kh a_\Kh \ ,\qquad
N^\Kh= \tfrac{1}{2} \int \Omega_c\wedge \beta^\Kh =
\tfrac{1}{2}\big(\xi^\Kh + 2i \R (C Z^\Kh)\big)\ . 
\eeq
Due to the orientifold projection the two three-forms 
$\Omega$ and $C_3$
each lost half of their degrees of freedom and combined
into a new complex three-form $\Omegac$. 
As we will show in more detail in the next section
the `good' chiral coordinates in the $N=1$ orientifold
are the periods of $C\Omega$ directly while in $N=2$
the periods agree with the proper field variables only 
in special coordinates.

Let us summarize the resulting $N=1$ spectrum.
It assembles into a gravitational multiplet,
$h^{(1,1)}_+$ vector multiplets and 
$(h^{(1,1)}_- + h^{(2,1)}+1)$ chiral multiplets.
We list the bosonic parts of the $N=1$ supermultiplets in table
\ref{N=1spectrum} \cite{BH}. We see that the $h^{(1,1)}$ $N=2$
vector multiplets split into $h^{(1,1)}_+$ $N=1$ vector multiplets and 
$h^{(1,1)}_-$ chiral multiplets while the 
$h^{(2,1)}+1$ hypermultiplets are reduced to $h^{(2,1)}+1$ chiral multiplets.

\begin{table}[h]
\begin{center}
\begin{tabular}{|l|c|c|} \hline 
 \rule[-0.3cm]{0cm}{0.8cm} 
multiplets& multiplicity & bosonic components\\ \hline\hline
 \rule[-0.3cm]{0cm}{0.8cm} 
 gravity multiplet&1&$g_{\mu \nu} $ \\ \hline
 \rule[-0.3cm]{0cm}{0.8cm} 
 vector multiplets&   $h_+^{(1,1)}$&  $A^{\alpha} $\\ \hline
 \rule[-0.3cm]{0cm}{0.8cm} 
 {chiral multiplets} &   $h_-^{(1,1)}$& 
$t^a$ \\ \hline
 \rule[-0.3cm]{0cm}{0.8cm} 
{chiral multiplets} 
& $ h^{(2,1)}+1$ &$ N^\Kh$\\ 
\hline
\end{tabular} 
\caption{\label{N=1spectrum} \textit{$N =1$ spectrum of orientifold compactification.}}
\end{center}
\end{table}

%
%

\subsection{The effective action \label{eff_act}}

In this section we calculate the four-dimensional effective action of type 
IIA orientifolds by performing a Kaluza-Klein reduction of the 
ten-dimensional type IIA action \eqref{10dact} taking the 
orientifold constraints into account. Equivalently this amounts to
imposing the orientifold projections on the $N=2$ action of 
section~\ref{revIIA}.
Inserting \eqref{expJB}, \eqref{decompO}, \eqref{form-exp} into
the ten-dimensional type IIA action \eqref{10dact} and performing a Weyl 
rescaling of the four-dimensional metric 
we find 
\bea \label{act1}
  S^{(4)}_{O6} &=& \int -\tfrac{1}{2} R*\mathbf{1} 
- G_{a b}\, dt^a \wedge * d \bar t^b 
+ \tfrac{1}{2} \text{Im}\, \cN_{\alpha \beta}\ F^\alpha \wedge * F^\beta 
        + \tfrac{1}{2} \text{Re}\, \cN_{\alpha \beta}\ F^\alpha
        \wedge F^\beta   \nn\\
      && 
\quad      -\, d D \wedge * dD  -\, G_{K L}(q)\, dq^K \wedge * dq^L 
         +\tfrac{1}{2} e^{2D}\, \text{Im}\, \cM_{ \Kh  \Lh}\, 
         d\xi^{\Kh} \wedge * d\xi^{\Lh} \ ,
\eea
where $F^\alpha = dA^\alpha$.
Let us discuss  the different couplings appearing in \eqref{act1}
in turn. 
Apart from the standard Einstein-Hilbert term the first line arises
from the projection of the $N=2$ vector multiplets action.
As we already observed the orientifold projection reduces the number
of K\"ahler moduli from $h^{(1,1)}$  to $h^{(1,1)}_-$ ($t^A\to t^a$)
but leaves the complex structure on this component of the moduli space
intact.  Accordingly the metric $G_{ab}(t)$ is inherited from the
metric $G_{AB}$ of the $N=2$ moduli space 
$\cM^{SK}$ given in \eqref{Kmetric}.
Since the volume form is odd only intersection numbers with one or
three odd basis elements in 
\eqref{int-numbers} can be non-zero and consequently one has
\beq \label{van-int}
  \cK_{\alpha \beta \gamma} = \cK_{\alpha a b} = \cK_{\alpha a} = \cK_{\alpha} = 0\ ,
\eeq  
while all other intersection numbers can be non-vanishing.\footnote{From a supergravity 
point of view this has been discussed also in \cite{ADAF}.} 
This implies that the metric $G_{AB}(t^A)$ of \eqref{Kmetric} is block
diagonal and obeys
\begin{eqnarray} \label{splitmetr}
  G_{a b}=
  -\frac{3}{2}\left( \frac{\KK_{a b}}{\KK}-
  \frac{3}{2}\frac{\KK_a \KK_b}{\KK^2} \right)\ , \qquad
  G_{\alpha \beta}=-\frac{3}{2} \frac{\KK_{\alpha \beta}}{\KK}\ , \qquad
G_{\alpha b}\ =\ 0\ ,
\end{eqnarray}
where
\begin{equation}\label{intO6}
  \KK_{ab}=\KK_{abc}\; v^c\ , 
\quad \ 
\KK_{\alpha \beta}=\KK_{\alpha \beta a}\; v^a\ ,\quad
\KK_{a}=\KK_{a b c}\; v^b v^c\ ,
  \quad \KK=\KK_{abc} \; v^a v^b v^c
\ .
\end{equation}

The same consideration also truncates  the $N=2$ gauge-kinetic 
coupling matrix $\cN_{\Ah \Bh}$ explicitly given in \eqref{def-cN}. 
Inserting \eqref{van-int} and \eqref{intO6} one arrives at 
\beq \label{def-N_alph_bet} 
  \text{Re} \cN_{\alpha \beta} = - \cK_{\alpha \beta a} b^a\ , \qquad 
  \text{Im} \cN_{\alpha \beta} = \cK_{\alpha \beta}\ ,\qquad \cN_{a \alpha}=\cN_{0 \alpha}=0\ .
\eeq
(The other non-vanishing matrix elements $\cN_{\ah\bh}$ arise in the potential 
\eqref{U-pot} once fluxes are turned on.) 

Let us now discuss the 
terms in the second line of \eqref{act1} arising from the
reduction  of the $N=2$ hypermultiplet action which is 
determined by the quaternionic metric \eqref{q-metr}.
$D$ is the the four-dimensional dilaton defined in \eqref{4d-dilaton}.
The metric $G_{KL}$ is inherited from the $N=2$
K\"ahler metric $G_{K \bar L}(z,\bar z)$ given in \eqref{csmetric}
and thus is 
the induced metric on the submanifold $\cM^{\rm cs}_\bbR$ 
defined by the constraint \eqref{constrO}.
More precisely, the complex structure deformations respecting \eqref{constrO}
can be determined from \eqref{Kod-form}
by considering infinitesimal variations of 
$\Omega$
\beq
  \Omega(z + \delta z) \ =\ \Omega(z) + \delta z^K (\partial_{z^K} \Omega)_{z} \ 
              =\ \Omega(z) - \delta z^K( \Kcs_{z^K} \Omega - \chi_K)_z \ .
\eeq 
Now we impose the condition that both 
$\Omega(z+\delta z)$ and $\Omega(z)$ satisfy \eqref{constrO}. 
This implies locally 
\beq  \label{constr2}
  \delta z^K\, \partial_{z^K} \Kcs = \delta\bar z^K\, \partial_{\bar z^K} \Kcs\ , \qquad 
  \delta z^K\sigma^* \chi_K = e^{2i \theta} \delta \bar z^K\bar \chi_K \ ,
\eeq 
where $\partial_{z^K} \Kcs$ and $\chi_K$ are restricted to $\cM^{\rm cs}_\bbR$.
Using the fact that $\Kcs$ is a K\"ahler potential and therefore $\partial_{z^K}\Kcs\neq 0$, we conclude from
the first equation in \eqref{constr2} that for each $\delta z^K$ either the 
real or imaginary part has to be zero. This is consistent with the observation
of the previous section that coordinates of $\cM^{\rm cs}_\bbR$ can be 
identified with 
the real or imaginary part of the complex structure deformations $z^K$.
To simplify the notation we call these deformations collectively
$q^K$ and denote the embedding map by 
$\rho:\cM^{\rm cs}_\bbR \hookrightarrow \cM^{\rm cs}$.
Locally this corresponds to 
\beq \label{embmap1}
  \rho:\ q^K=(q^s,q^\sigma)\ \mapsto\ z^K=(q^s,iq^\sigma)\ ,
\eeq
for some splitting $z^K=(z^s,z^\sigma)$. In other words, 
the local coordinates on $\cM^{\rm cs}_\bbR$ 
are $\R z^s=q^s$ and $\I z^\sigma = q^\sigma$ while $\I z^s=0=\R z^\sigma$.
Using the second equation in \eqref{constr2}, the embedding
map \eqref{embmap1} and the expression \eqref{chi_barchi} for the $N=2$ metric $G_{K\bar L}$ we also deduce that
the K\"ahler form vanishes when pulled back to $\cM^{\rm cs}_\bbR$. 
In summary we 
have
\beq \label{def-G}
  \rho^*(G_{K \bar L}\, dz^K d \bar z^L)\, \equiv\, G_{KL}(q)\, dq^K dq^L\ , \quad
  \rho^*(iG_{K \bar L}\, dz^K \wedge d \bar z^L)\, =\, 0\ .
\eeq
The first equation defines the induced metric while the second equation
implies that $\cM^{\rm cs}_\bbR$
is a Lagrangian submanifold of $\cM^{\rm cs}$ with respect to the 
K\"ahler-form.

Finally, coming back to the action \eqref{act1}
the matrix $\cM_{\Kh \Lh}$ is defined in analogy with
\eqref{defM} as
\bea \label{defM2}
  \int a_\Kh \wedge * a_\Lh&=&-\text{Im}\; \cM_{\Kh \Lh} 
\ , \qquad\qquad
\int a_\Kh\wedge * b^\Lh = 0\ , \nn\\
 \int b^\Kh \wedge * b^\Lh\ &=&\ -(\text{Im}\; \cM)^{-1\ \Kh \Lh}\ , 
\eea
where 
$\text{Im} \cM_{\Kh \Lh}$ can be given explicitly 
in terms of the periods by inserting \eqref{Z=0} into \eqref{gauge-c}.
This yields
\bea \label{gauge-r}
\I \cM_{\Kh \Lh}=-\I \mathcal{F}_{\Kh \Lh}+2 \frac{(\text{Im}\; \mathcal{F})_{\Kh \Mh} \R (CZ^\Mh)
   (\text{Im}\; \mathcal{F})_{\Lh \Nh}\R (CZ^\Nh) }{\R (CZ^\Nh)(\text{Im}\; \mathcal{F})_{\Nh\Mh} 
    \R (CZ^\Mh)}\ .
\eea 
Similarly one obtains $\R\cM_{\Kh \Lh}=0$ 
consistent with \eqref{defM} which corresponds to the 
vanishing of the second intersection in \eqref{defM2}.

This ends our discussion of
the effective action obtained by applying the orientifold projection. 
The next step is to rewrite the action \eqref{act1} 
in the standard $N=1$ supergravity form
which we turn to now.

\subsection{The effective action in the N=1 supergravity form \label{eff_supform}} 

In 
$N=1$ supergravity the action is expressed 
 in terms of a K\"ahler potential $K$, 
a holomorphic superpotential $W$ and the holomorphic gauge-kinetic coupling 
functions $f$ as follows \cite{WB,GGRS}
\beq\label{N=1action}
  S^{(4)} = -\int \tfrac{1}{2}R * \mathbf{1} +
  K_{I \bar J} dM^I \wedge * d\bar M^{\bar J}  
  + \tfrac{1}{2}\text{Re}f_{\alpha \beta}\ 
  F^{\alpha} \wedge * F^{\beta}  
  + \tfrac{1}{2}\text{Im} f_{\alpha \beta}\ 
  F^{\alpha} \wedge F^{\beta} + V*\mathbf{1}\ ,
\eeq
where
\beq\label{N=1pot}
V=
e^K \big( K^{I\bar J} D_I W {D_{\bar J} \bar W}-3|W|^2 \big)
+\tfrac{1}{2}\, 
(\text{Re}\; f)^{-1\ \alpha\beta} D_{\alpha} D_{\beta}
\ .
\eeq
Here the $M^I$ collectively denote all
complex scalars in the theory  and 
$K_{I \bar J}$ is a K\"ahler metric satisfying
$  K_{I\bar J} = \partial_I \bar\partial_{\bar J} K$.
The scalar potential is expressed in terms of the 
K\"ahler-covariant derivative $D_I W= \partial_I W + 
(\partial_I K) W$. 

Comparing \eqref{act1} with \eqref{N=1action} 
using \eqref{def-N_alph_bet} and \eqref{def-t}
we can immediately
read off the gauge-kinetic coupling function $f_{\alpha \beta}$ to be
\beq \label{gauge-A}
  f_{\alpha \beta}\ =\ -i \bar \cN_{\alpha \beta}\ =\ i \cK_{\alpha \beta a}  t^a \ .
\eeq
As required by $N=1$ supersymmetry the $f_{\alpha \beta}$ 
are indeed holomorphic. Note that they are linear in the $t^a$ moduli
and do not depend on the complex structure and $\xi$-moduli.

{}From \eqref{act1} we also immediately observe that
the orientifold moduli space has the product structure  
\beq \label{direct-mod}
  \tilde\cM^{\rm K} \times \tilde\cM^{\rm Q}\ .
\eeq
The first factor $\tilde\cM^{\rm K}$ is a subspace of the 
$N=2$ moduli space $\cM^{\rm K}$ with dimension $h^{(1,1)}_-$ 
spanned by  the complexified K\"ahler deformations $t^a$.
The second factor $\tilde\cM^{\rm Q}$ is a subspace of the quaternionic
manifold $\cM^{\rm Q}$ with
dimension $h^{(2,1)} +1$ 
spanned by  the complex structure deformations $q^K$, the dilaton $D$
and the scalars $\xi^{\hat K}$ arising from $C_3$.
Let us discuss both factors in turn.

As we already stressed earlier the metric $G_{ab}$ 
of \eqref{act1} defined in \eqref{splitmetr} is a trivial
truncation
of the $N=2$ special K\"ahler metric  \eqref{Kmetric} and therefore remains
special K\"ahler. The K\"ahler potential is given by
\beq \label{Kks}
K^{\rm K}\ =\ - \ln \Big[\tfrac{i}{6}\cK_{a b c} (t -\bar t)^a (t -\bar t)^b (t -\bar t)^c \Big]
  \ = \ - \ln \Big[\tfrac{4}{3}  \int_Y J \wedge J \wedge J\Big]\ ,
\eeq
where $J$ is the K\"ahler form in the string frame.
Moreover, $K^{\rm K}$ can be obtained from 
the prepotential $f(t)=-\tfrac{1}{6} \cK_{abc}t^a t^b t^c$ 
by using equation \eqref{Kinz}. 
It is well known that $K^{\rm K}$ obeys the standard no-scale condition
\cite{NS}
\beq \label{no-scale1}
  K_{t^a} K^{t^a \bar t^b} K_{ \bar t^b}\ =\ 3\ . 
\eeq

The geometry of the second component  $\tilde\cM^{\rm Q}$ in \eqref{direct-mod}
is considerably more complicated. This is due to the fact that 
\eqref{newO} defines a new complex structure on the field space. In the following
we sketch the calculation of the K\"ahler potential for the basis 
$(a_\Kh,b^\Kh)$ and only summarize the results for a generic symplectic basis.
The details of this more involved calculation will be presented in appendix \ref{Geom-of-M}. 

To begin with, let us define the compensator $C$ introduced in section \ref{spectrum} as 
\beq \label{def-C}
   C\ =\ e^{-D-i\theta} e^{\Kcs(q)/2}\ , \qquad C \rightarrow C e^{\R\, h(q)}\ ,
\eeq 
where $\Kcs$ is the K\"ahler potential defined in \eqref{csmetric} restricted
to the real subspace $\cM^{\rm cs}_{\bbR}$. We also displayed the transformation 
behavior of $C$ under real K\"ahler transformations \eqref{real_K}. With this at hand one 
defines the scale invariant variable
\beq \label{l-def}
   l^\Kh \ =\  \R(C Z^\Kh(q))\ .
\eeq
Inserted into \eqref{act1} and using 
the Jacobian matrix encoding the change of variables $(e^D,q^K) \rightarrow l^\Kh$
the second line \eqref{act1} simplifies as\footnote{%
The calculation of this result can be found in appendix \ref{Geom-of-M}.}
\beq \label{IIAQ}
  \cL^{(4)}_{\rm Q} =  2 e^{2D}\, \text{Im}\, \cM_{\Kh \Lh}\, 
(dl^\Kh \wedge * dl^\Lh  + \tfrac{1}{4} d\xi^{\Kh} \wedge * d\xi^{\Lh})\ .  
\eeq
We see that the scalars $l^\Kh$ and $\xi^\Kh$ nicely combine 
into complex coordinates 
\beq \label{Ncoords}
   N^\Kh\ =\ \tfrac{1}{2}\xi^\Kh +  i l^\Kh\ 
=\ \tfrac{1}{2}\xi^\Kh +  i \R(C Z^\Kh)
= \tfrac{1}{2} \int \Omega_c\wedge b^\Kh 
\ , 
\eeq
which we anticipated in equation \eqref{newO}.
The important fact to note here is that $\tilde\cM^{\rm Q}$
is equipped with a new complex structure and the corresponding
K\"ahler coordinates 
coincide with half of the periods of $\Omegac$.
This is in contrast to the situation in $N=2$ where one of the periods
($Z^0$) is a gauge degree of freedom and the K\"ahler
coordinates are the special coordinates $z^K = Z^K/Z^0$.

In order to show that the metric in \eqref{IIAQ} is K\"ahler we need the
explicit expression for the K\"ahler potential. Using \eqref{gauge-r} one 
obtains straightforwardly
\beq
2 e^{2D} \text{Im}\, \cM_{\Kh \Lh} = \partial_{N^\Kh} \partial_{\bar N^\Lh} K^{\rm Q}\ ,
\eeq
where 
\beq \label{KQsimple}
 K^{\rm Q} = -2 \ln\big[4i\cF(CZ)\big]\ , \qquad
\cF\big(\R(CZ)\big) = \frac{i}2\,  \R(C Z^\Kh)\, \I(C\cF_\Kh)\ .
\eeq
Alternatively, using \eqref{decompO} and $*\Omega =- i \Omega$
one derives the integral representation
\beq \label{intKQ}
  K^{\rm Q}\ = - 2\ln\Big[2\int_Y \R(C \Omega)\wedge *\R(C\Omega)\Big]=\ - \ln\, e^{-4D} \ , 
\eeq
where in the second equation we used \eqref{def-C} and \eqref{csmetric}. In 
the form \eqref{intKQ} the dependence of $K^{\rm Q}$ on the coordinates $N^\Kh$
is only implicit and given by means of their definition \eqref{Ncoords}.  
Also $K^{\rm Q}$ obeys a no-scale type condition in that it 
satisfies
\bea \label{no-scale2}
  K_{N^\Kh} K^{N^\Kh \bar N^\Lh} K_{\bar N^\Lh} = 4\ ,
\eea
which can be checked by direct calculation.

The analysis so far started from the symplectic basis 
$(a_\Kh,b^\Kh)$ introduced in \eqref{basis_ab}, 
determined the K\"ahler coordinates in \eqref{Ncoords}
and derived  the K\"ahler potential $K^{\rm Q}$
in terms of the prepotential $\cF$ in \eqref{KQsimple} or as an 
integral representation in \eqref{intKQ}. Now we need to ask
to what extent  this result depends on the choice of 
the basis \eqref{basis_ab}. Or in other words let us redo
the calculation starting from an arbitrary symplectic basis
and determine the K\"ahler potential and the proper field variables
for the corresponding orientifold theory. 
Let us first recall the situation 
in the $N=2$ theory reviewed in section \ref{revIIA}. 
The periods $(Z^\Kh,\cF_\Kh)$ defined in \eqref{pre-z}
form a symplectic vector
of $Sp(2h^{(1,2)}+2,\bf Z)$
such that $\Omega$ given in \eqref{Omegaexp} and 
$\Kcs$ given in \eqref{csmetric} is manifestly invariant.
The prepotential  $\cF(Z) = \frac{1}{2} Z^\Kh \cF_\Kh$ on the other hand 
does depend
on the choice of the basis $(\alpha_\Kh,\beta^\Kh)$
and is not invariant.  

For $N=1$ orientifolds this situation is different 
since the orientifold projection \eqref{constrO} explicitly breaks the 
symplectic invariance.\footnote{A symplectic transformation $\cS$ preserve the
form $\big<\alpha,\beta\big> = \int \alpha \wedge \beta$, such that 
$\big<\cS \alpha,\cS \beta \big> = \big< \alpha,\beta \big>$.
On the other hand the anti-holomorphic involution satisfies
$\big<\sigma^* \alpha,\sigma^* \beta \big>  
= - \big< \alpha,\beta \big>$.}  
This can also be seen from the form
of the $N=1$ K\"ahler potential \eqref{KQsimple} which is expressed
in terms of the non-invariant prepotential.
One immediately concludes that the result \eqref{KQsimple} is 
basis dependent and $K^Q$ takes this simple form due to the special 
choice $a_\Kh \in H^{3}_+(Y)$ and $b^\Kh \in H^3_-(Y)$.\footnote{Note that this is in striking analogy to 
the background dependence of the B model partition function as discussed in \cite{BCOV,Witten2}}
On the other hand, the integral representation \eqref{intKQ} only implicitly depends 
on the symplectic basis through the definition of the coordinates $N^\Kh$. 
This suggest, that it is possible to generalize our results by allowing for 
an arbitrary choice of symplectic basis in the definition of the $N=1$ coordinates. 
More precisely, let us consider the generic basis $(\alpha_\Kh,\beta^\Lh)$, 
where we assume that the $h^3_+=h^{2,1}+1$ basis elements $(\alpha_k,\beta^\lambda)$
span $H^3_+$ and the $h^3_-=h^{2,1}+1$ basis elements $(\alpha_\lambda,\beta^k)$ span $H^3_-$.
In this basis the intersections \eqref{symplectic} take the form
\beq \label{sp_alpha-beta}
  \int_Y \alpha_k \wedge \beta^l\ =\ \delta_k^l\ , \qquad 
  \int_Y \alpha_\kappa \wedge \beta^\lambda\ =\ \delta_\kappa^\lambda\ ,
\eeq
with all other combinations vanishing.
Applying the orientifold constraint \eqref{constrO} one concludes that 
the equations \eqref{Z=0} are replaced by 
\beq \label{Z=0gen}
  \I(C Z^k) =  \R (C \cF_k )\ =\ 0\ , \qquad 
   \R (C Z^\lambda) = \I(C \cF_\lambda)\ =\ 0\ .
\eeq 
Correspondingly, the expansions \eqref{decompO} and \eqref{form-exp}
take the form
\bea \label{decompO2}
  C \Omega &=& \R (C Z^k) \alpha_k + i\I (C Z^\lambda ) \alpha_\lambda -
             \R (C \cF_\lambda) \beta^\lambda - i\I (C \cF_k) \beta^k\ ,\nn\\
\CC_3 &=& \xi^k\, \alpha_k - \tilde \xi_\lambda\, \beta^\lambda\ ,
\eea
which implies that we also have to redefine the $N=1$ coordinates of 
$\tilde \cM^{\rm Q}$ in an appropriate way. 
In appendix~\ref{Geom-of-M} we show that the 
new K\"ahler coordinates  $(N^k,T_\lambda)$ are again determined by the periods of $\Omegac$ and given  by 
\bea \label{Oexp}\label{def-NT}
  N^k &=&\tfrac{1}{2} \int \Omegac \wedge \beta^k \
 =\ \tfrac{1}{2}\xi^k + i \R(CZ^k)\ , \nn\\
  T_\lambda &=& i \int \Omegac \wedge \alpha_\lambda\ =\
i\tilde \xi_\lambda - 2 \R (C \cF_\lambda) \ ,
\eea
where we evaluated the integrals by using \eqref{Omegacdef} 
and \eqref{decompO2}.

The K\"ahler potential takes again the form \eqref{intKQ} but now 
depends on $N^k,T_\lambda$ and thus no longer simplifies to \eqref{KQsimple}.
Let us compare the situation to the original $N=2$ theory, which
was formulated in terms of the
$Z^\Kh$ or equivalently the special coordinates $z^K$. Holomorphicity
in these coordinates played a central role in defining the prepotential
encoding the special geometry of $\cM^{\rm cs}$ in $\cM^{\rm Q}$ (cf.~section 
\ref{revIIA}). In contrast, the $N=1$ orientifold constraints destroy this complex structure and force us 
to combine $\R(C\Omega)$ with the RR three-form $C_3$ into $\Omegac$. The 
K\"ahler coordinates are half of the periods of $\Omegac$ 
but now in this more general case also the
derivatives of $\cF$ can serve as coordinates as seen in \eqref{def-NT}. 
However, as it is shown in appendix 
\ref{Geom-of-M}, $\R (C \cF_\lambda)$ and $e^{2D}\I (CZ^\lambda)$ are related
by a Legendre transformation of the K\"ahler potential. Working with this transformed
potential and the coordinates $\R(CZ^k)$ and $e^{2D}\I (CZ^\lambda)$ enables us 
to make contact to the underlying $N=2$ theory in its canonical formulation. 
From a supergravity point of 
view, this Legendre transformation corresponds to replacing the chiral multiplets 
$T_\lambda$ by linear multiplets as described in appendix \ref{linm} and \ref{Geom-of-M}.
This is possible due to the translational isometries of $K$, 
which arise as a consequence of the $C_3$ gauge invariance 
and which render $K$ independent of
the scalars $\xi$ and $\tilde\xi$. 
We show in appendix \ref{Geom_modspace}
that this also enables us to construct $\tilde \cM^{\rm Q}$ from $\cM^{\rm cs}_\bbR$
similar to the moduli space of supersymmetric Lagrangian submanifolds in a
Calabi-Yau space as described by Hitchin \cite{Hitchin2}.  
This also allows us to interpret the no-scale condition \eqref{no-scale2}
geometrically.

Let us summarize the results obtained so far. We found that the moduli
space of $N=1$ orientifolds is indeed the product of two K\"ahler spaces
with the K\"ahler potential 
\beq \label{N=1Kpot}
  K\ =\ K^{\rm K} + K^{\rm Q} = - \ln \Big[\tfrac{4}{3}  \int_Y J \wedge J \wedge J\Big] 
        - 2\ln\Big[2\int_Y \R(C \Omega)\wedge *\R(C\Omega)\Big]\ .
\eeq 
The first term depends on the K\"ahler deformations of the orientifold
while the second term is a function of the real complex structure 
deformations and the dilaton.
The $N=1$ K\"ahler coordinates are obtained 
by expanding the complex combinations\footnote{This combination 
of forms has also appeared recently in ref.\ \cite{NOV}
in the discussion of $D$-instanton couplings in the A-model.
Here they appear as the proper chiral $N=1$ variables and as we will
see in the next section they linearize the D-instanton action.}
\beq \label{N=1coords}
  \Omegac\ =\ \CC_3 + 2i \R(C\Omega)\ ,\qquad 
  \Jc\ =\ \hat B_2 + iJ \ , 
\eeq
in a real harmonic basis of $H^{3}_+(Y)$ and $H^{(1,1)}_-(Y)$ respectively. 
Note that $K$ does not depend on the scalars arising in the expansion of 
$\hat B_2$ and $\hat C_3$, such that the K\"ahler manifold admits a set of 
$h^{(1,1)}_- + h^{(2,1)}+1$ translational isometries. In other words
$K$ consists of two functionals encoding the dynamics of the two-form $J$ 
and the real three-form $\R(C\Omega)$.\footnote{The functions 
$V [ \R(C\Omega) ]= \int \R(C\Omega)\wedge* \R(C\Omega)$
and $V [J ]=\int J\wedge J \wedge J$ are known as Hitchins functionals \cite{Hitchin1}. 
The orientifold constraints \eqref{constrJ} and \eqref{constrO} 
restricts their domain to $J \in H^2_-(Y)$
and $\R(C\Omega) \in H^3_+(Y)$.}   
Moreover, irrespective of the chosen basis the K\"ahler potential 
obeys the no-scale type conditions \eqref{no-scale1} and 
\eqref{no-scale2}, \eqref{no-scale4}.

However, these two statements are violated when further stringy 
corrections are included. $K$ receives additional contributions due
to perturbative effects as well as world-sheet and $D2$ instantons.
It is well-known that the combination $\Jc=\hat B_2 + i J$ 
gives the proper coupling to the string world-sheet such that 
 world-sheet instantons correct the holomorphic prepotential as 
$f(t) = -\frac{1}{6}\cK_{abc}t^a t^b t^c + O(e^{-t})$. 
Since we divided out the world-sheet parity these corrections also
include non-orientable Riemann surfaces, such that the prepotential 
$f(t)$ consists of two parts $f(t) = f_{or}(t) + f_{unor}(t)$. 
The function $f_{or}$ counts holomorphic maps 
from orientable world-sheets to $Y$, while $f_{unor}$ counts holomorphic maps 
from non-orientable world-sheets to $Y$ \cite{BFM}. 
In the
next section we show that $D2$ instantons naturally couple to the complex three-form  $\Omegac$ and they are expected to correct 
$K^{\rm Q}$.


\section{The effective action in the presence of background fluxes \label{fluxes_sec}}
In this section we derive the effective action of type IIA orientifolds
in the presence of background fluxes. 
For standard $N=2$ Calabi-Yau compactifications of type IIA a
similar analysis is carried out in refs.\ \cite{LM,KachruK}.
In order to do so
we need to start from the ten-dimensional action of massive 
type IIA supergravity which differs from the action \eqref{10dact} in
that the two-form $\hat B_2$ is massive. In the 
Einstein frame it is given by \cite{Romans}
\bea \label{10dactm}
  S^{(10)}_{MIIA} &=& \int -\tfrac{1}{2}\hat R*\mathbf{1} -\tfrac{1}{4} d\hat \phi\wedge * d\hat \phi
  -\tfrac{1}{4} e^{-\hat \phi}\hat H_3 \wedge *\hat H_3 
  -\tfrac{1}{2} e^{\frac{3}{2} \hat \phi}\hat F_2 \wedge *\hat F_2 \nn \\
  && -\tfrac{1}{2} e^{\frac{1}{2} \hat \phi}\hat F_4 \wedge *\hat F_4 
  -\tfrac{1}{2} e^{\frac{5}{2} \hat \phi}\, (m^0)^2 * \mathbf{1} + \cL_{\rm top}\ ,
\eea
where
\bea
  \cL_{\rm top}&=& -\tfrac{1}{2}\Big[ \hat B_2 \wedge d\hat C_3 \wedge d\hat C_3\  
                                   -(\hat B_2)^2 \wedge d\hat C_3 \wedge d\hat A_1
                                   + \tfrac{1}{3}(\hat B_2)^3 \wedge (d\hat A_1)^2 \nn \\
               & &                 - \tfrac{m^0}{3}(\hat B_2)^3 \wedge d\hat C_3
                                   + \tfrac{m^0}{4}(\hat B_2)^4 \wedge d\hat A_1
                                   + \tfrac{(m^0)^2}{20}(\hat B_2)^5 \Big]\ ,
\eea
and the field strengths are defined as
\bea \label{defHFF}
  \hat H_3 = d \hat B_2\ , \quad \hat F_2 = d\hat A_1+m^0 \hat B_2\ , \quad 
  \hat F_4 = d\hat C_3 - \hat A_1 \wedge \hat H_3-\tfrac{m^0}{2}(\hat B_2)^2\ .
\eea
Compared to the analysis of the previous section we now include
non-trivial background fluxes of the field strengths
$F_2$, $H_3$ and $F_4$ on the Calabi-Yau orientifold.
We keep the Bianchi identity and the equation of motion intact 
and therefore expand $F_2$, $H_3$ and $F_4$
in terms of  harmonic forms compatible with the orientifold
projection. From \eqref{fieldtransf} we infer that $F_2$ is expanded in
harmonic forms of $H^{2}_-(Y)$, 
$H_3$ in harmonic forms of $H^3_{-}(Y)$ and $F_4$ in harmonic forms
of $H^{4}_+(Y)$.\footnote{As we observed in the previous section
there is no $\hat A_1$
due to the absence of one-forms on the orientifold. 
Nevertheless its field strength $F_2$ 
can be non-trivial on the orientifold since $Y$ generically possesses
non-vanishing harmonic two-forms.}
 Explicitly the expansions read 
\bea \label{fluxes}
 H_3\, =\, q^\lambda \alpha_\lambda - p_k\, \beta^k\ , \quad   F_2\, =\, -m^a \omega_a\ , \quad 
  F_4\, =\, e_a\, \tilde \omega^a\ ,
\eea
where $(q^\lambda,p_k)$ are $h^{(2,1)}+1$ real NS flux parameters 
while $(e_a,m^a)$ are $2h^{1,1}_-$ real RR flux parameters.
The harmonic forms $(\alpha_\lambda, \beta^k)$ are the elements of the real
symplectic basis of $H^3_-$ introduced in \eqref{sp_alpha-beta}. 
The basis $\tilde \omega^a$ of
$H^{(2,2)}_+$ is defined to be the dual basis of $\omega_a$ while the
basis $\tilde \omega^\alpha$ denotes a basis of $H^{(2,2)}_-$ dual to $\omega_\alpha$. 

Inserting \eqref{expJB}, \eqref{form-exp} and \eqref{fluxes} into
\eqref{defHFF} we arrive at
\bea \label{fieldst}
  \hat H_3 &=& db^a\wedge \omega_a + q^\lambda \alpha_\lambda - p_k\, \beta^k\ ,  \qquad \qquad 
  \hat F_2 = (m^0 b^a + m^a)\, \omega_a\ ,\\
  \hat F_4 &=& dC_3 + dA^\alpha \wedge \omega_\alpha 
  + d\xi^k \wedge \alpha_k - 
            d\tilde \xi_\lambda \wedge \beta^\lambda +  
   \big(b^a m^b  -\tfrac12 m^0 b^a b^b\big)\, 
\cK_{abc}\tilde \omega^c + e_a\, \tilde \omega^a\ , \nn
\eea
where we have used $\omega_a \wedge \omega_b = \cK_{abc}\, \tilde \omega^c$.
Now we repeat the KK-reduction of the previous section using the
modified field strength 
\eqref{fieldst} and the action \eqref{10dactm} instead of \eqref{10dact}.
This results in%
\footnote{The action $S^{(4)}_{O6}$ is given in \eqref{act1}. However, due to the fact that 
we perform the Kaluza-Klein reduction in the generic basis introduced in \eqref{sp_alpha-beta} the kinetic 
terms for $\tilde \cM^{\rm Q}$ are replaced by \eqref{act2}. }
\beq\label{Sflux}
S^{(4)} = S^{(4)}_{O6} - \int  \tfrac{g}{2}\, d\cc_3 \wedge * d\cc_3 + {h}\, d\cc_3 +
 U * \mathbf{1}\ ,
\eeq
where $S^{(4)}_{O6}$ is given in \eqref{act1}.
$\cc_3$ 
is the four-dimensional part of the ten-dimensional 
three-form $\hat C_3$ defined in \eqref{form-exp} and 
its couplings to the scalar fields are given by
\beq
  g =  e^{-4 \phi} \left(\tfrac{\cK}6
       \right)^3\ , \qquad  h = e_a b^a + \tilde \xi_\lambda q^\lambda 
  - \xi^k p_k + \tfrac{1}{2}\R \cN_{0 \ah}\, m^\ah \ ,
\eeq
where we denoted $m^\ah=(m^0,m^a)$. The potential term $U$ of \eqref{Sflux}
is given by
\beq \label{U-pot}
  U = \tfrac{9}{\cK^2} e^{2\phi} \int_Y H_3 \wedge * H_3
        -  \tfrac{18}{\cK^2} e^{4\phi} \I \cN_{\ah \bh}\, m^\ah m^\bh
         +\tfrac{ 27 } {\cK^3} e^{4\phi} G^{ab}(e_a - \R \cN_{a\ah}\, m^\ah)(e_b - \R \cN_{b\bh}\, m^\bh), 
\eeq
where
\beq
  \int_Y H_3 \wedge * H_3 = -  (p_k - \R \cM_{k \lambda} q^\lambda)(\I \cM)^{-1\, k l}
                 (p_l - \R \cM_{l \lambda} q^\lambda) - \I \cM_{\kappa \lambda}\, q^\kappa\, q^\lambda\ .
\eeq
The matrix $\cN_{\ah \bh} (t,\bar t)$ is defined to be the corresponding 
part of the $N=2$ gauge-coupling matrix \eqref{def-cN} 
restricted to $\tilde \cM^{\rm K}$ by applying \eqref{van-int} and \eqref{splitmetr}. Similarly the matrices 
$\cM_{l \lambda}, \cM_{\kappa \lambda}, \cM_{k l}$ are obtained from
 the $N=2$ matrix $\cM_{\Kh\Lh}$ defined in  \eqref{gauge-c}
by applying the orientifold constraints \eqref{Z=0gen},
i.e.\ restricting them to the subspace $\cM_\bbR^{\rm cs}$.

In four space-time dimensions 
$\cc_3$ is dual to constant which plays the role of
an additional electric flux $e_0$ in complete analogy with the
situation in $N=2$ discussed in \cite{LM}.
In order to write the 
action in terms of $e_0$ instead of $\cc_3$ we follow \cite{LM}
and add it as a Lagrange multiplier to the action \eqref{Sflux} 
\beq \label{S_C3}
S^{(4)} \to S^{(4)} + {e_0}\, d\cc_3 \ .
\eeq
Treating $d\cc_3$ as an independent four-form its 
equation of motion reads $*d\cc_3 = - (h+e_0)/g$ which can be used
to eliminate $d\cc_3$ in favor of $e_0$.\footnote{An alternative 
derivation is given in ref.\ \cite{BW}. Minimizing 
$U$ with respect to $d\cc_3$
 sets it to the value $*d\cc_3 = - h/g$.  Inserted 
back into $U$ only gives its classical value while quantum mechanical
states labeled by integers $e_0$ shift $h$ as given in \eqref{V+h}.} 
Inserted back into \eqref{S_C3} one 
finds
\beq \label{S_e_0}
 S^{(4)} = S^{(4)}_{O6} + \int V * \mathbf{1} \ ,
\eeq
where
\beq \label{V+h}
V = U + \int \frac{1}{2g}\, (h+e_0)^2\ .
\eeq
Inserting \eqref{U-pot} we arrive at
\bea \label{V-pot1}
   V & =& \tfrac{9}{\cK^2} e^{2\phi} \int H_3 \wedge * H_3 
      - \tfrac{18}{\cK^2} e^{4\phi} (\tilde e_\ah - \cN_{\ah \ch}\, m^\ch) (\I \cN)^{-1\, \ah \bh}
                                          (\tilde e_\bh -\bar \cN_{\bh
        \ch}\, m^\ch)\ ,
\eea
where we introduced the shorthand notation 
$\tilde e_\ah=(e_0 + \xi_\lambda q^\lambda-\xi^\kh p_\kh,e_a)$ and $m^\ah=(m^0,m^a)$. 
Note that in the presence of NS flux 
one can absorb $e_0$ by shifting the fields 
$\xi,\tilde \xi$. This corresponds to adding an integral form to 
$\CC_3$ as carefully discussed in \cite{BW}. 
However, for the discussion of mirror symmetry it is more convenient to
keep the parameter $e_0$ explicitly in the action.

In order to establish the consistency with $N=1$ supergravity
we need to rewrite $V$ given in \eqref{V-pot1} in terms of
\eqref{N=1pot}
or in other words we need express $V$ in terms of a superpotential 
$W$ and appropriate $D$-terms. From \eqref{Sflux} we infer that
turning on fluxes does not charge any of the fields and therefore
all $D$-terms have to vanish.\footnote{In type IIB 
orientifolds with $O5/O9$ planes a D-term and massive tensor fields
appeared when NS-flux are turned on \cite{GL}. 
The mirror symmetric situation corresponds to compactifications
on half-flat manifolds exactly as in $N=2$ \cite{GLMW}. Work along
these lines is in progress \cite{GLprep}.}
In order to determine $W$ we first need to compute the inverse
K\"ahler metric. Using \eqref{invKm1}, \eqref{lLmetric} and \eqref{defM} we find 
\bea \label{invM-TN}
  K^{T_\kappa \bar T_\lambda}& =& 2 e^{-2D} \int \alpha_\kappa \wedge * \alpha_\lambda\ , \qquad
  K^{T_\lambda \bar N^k}\ =\ i e^{-2D} \int \alpha_\lambda \wedge *
  \beta^k \ ,
\nn \\
  K^{N^k \bar N^l} & =& \tfrac12 e^{-2D} \int \beta^k\wedge * \beta^l
  \ , 
\qquad  
  K^{t^a \bar t^a} \ =\  G^{ab}\ . 
\eea
With the help of \eqref{invM-TN}, \eqref{N=1pot} and \eqref{N=1Kpot} one checks
that the potential \eqref{V-pot1} can be entirely expressed in
terms of the superpotential 
\beq \label{superpot1}
  W\ =\  W^{\rm Q}(N,T) + W^{\rm K}(t)\ ,
\eeq 
where 
\bea \label{superpot2}
  W^{\rm Q}(N^k,T_\lambda)& =& \int_Y \Omegac \wedge H_3\ =\ 
        - 2N^k p_k - i T_\lambda q^\lambda\ , \\
  W^{\rm K}(t^a) &=& e_0 + \int_Y \Jc \wedge F_4 - \tfrac{1}{2} \int_Y \Jc \wedge \Jc \wedge F_2 
       - \tfrac{1}{6} m^0 \int_Y \Jc \wedge \Jc \wedge \Jc\ ,
\nn\\
&=& e_0 + e_a t^a + \tfrac{1}{2}\cK _{abc} m^a t^bt^c - \tfrac{1}{6} m^0  \cK _{abc} t^a t^bt^c\nn\ ,
\eea
and $\Omegac$ and $\Jc$ are defined in \eqref{N=1coords}.
We see that the superpotential is the sum of two terms.
$W^{\rm Q}$  depends on the NS fluxes $(p_k,q^\lambda)$ of $H_3$ and the 
chiral fields $N^k,T_\lambda$ parameterizing the space $\tilde \cM^{\rm Q}$. 
$W^{\rm K}$ depends on the RR fluxes $(e_{\hat a}, m^{\hat b})$  
of $F_2$ and $F_4$ (together with $m^0$ and $e_0$) and
the complexified K\"ahler deformations $t^a$ parameterizing 
$\cM^{\rm K}$. 
We see that contrary to the type IIB case both types of moduli, 
K\"ahler and complex structure deformations appear in the superpotential
suggesting the possibility that all moduli can be fixed in this set-up.
This has recently also been observed in ref.~\cite{KachruK}.
A more detailed phenomenological investigation will be presented elsewhere.

Let us close this section by briefly discussing possible instanton
corrections
to the superpotential \eqref{superpot1}. They can arise from
worldsheet instantons 
wrapping the string around two-cycles of the orientifold 
or from wrapping $D2$-branes around three-cycles $\Sigma_3$ \cite{BBS}.  
The first set of corrections 
 contribute analogously to the  $N=2$ theory  with the difference
that also non-oriented worldsheets can contribute as discussed at the end
of the previous section.

The second set of correction comes from wrapping $D2$-branes around three-cycles
and can be viewed as the mirror symmetric corrections to the ones
discussed in \cite{Witten}. A computation of such corrections
is beyond the scope of this paper but let us make the observation that
they amount to holomorphic contribution in $W$
when expressed in the proper K\"ahler variables \eqref{def-NT}.
This can be seen from the fact that any correlation function
is weighted by the string-frame 
world-volume action of the wrapped Euclidean $D2$-branes
and thus includes a factor $e^{-S_{D2}}$ where\footnote{The possible
extra term $\hat A_1\wedge \hat B_2$  
does not appear in the  Chern-Simons part of \eqref{instact}
 since $\hat A_1$ is projected 
out by the orientifold.} 
\beq \label{instact}
  S_{D2} = -\mu_3\, e^{- \hat \phi} 
  \int_{\cW_3}d^3 \lambda \sqrt{\det\big({\varphi^*(\hat g+ \hat B_2) +  2\pi \alpha' F_2}\big)} 
             + i\mu_3 \int_{\cW_3} \varphi^*(\hat C_3)\ .
\eeq 
$\cW_3$ is the world-volume of the $D2$-brane and  $\varphi^*$ is the pullback
of the map $\varphi$ which embeds $\cW_3$ into Calabi-Yau orientifold $Y$ 
$\varphi:\cW_3 \hookrightarrow Y$.
The first term is the Dirac-Born-Infeld action describing the couplings 
of the $D2$-brane to the bulk metric and the bulk $\hat B$-field while 
the second term  is the Chern-Simons action which 
represents the coupling to the RR 3-form $\hat C_3$. 
We have chosen the RR charge $\mu_3$ equal to the tension since 
the wrapped  $D2$-branes must be BPS 
in order to preserve $N=1$ supersymmetry.
In fact there is an additional condition arising from 
the requirement that the $D2$-branes  preserves 
the same supersymmetry that is left intact
by the orientifold projections. This in turn implies 
that both the $D2$-brane and the internal part of the 
$O6$-planes wrap special Lagrangian cycles calibrated
with respect to the same real three-form.

The calibration condition for Euclidean $D2$-branes
has been derived in refs.\ \cite{BBS,MMMS}.
In order to adjust the normalization to the case at hand let us recall
that the unbroken supercharge 
has to be some linear combination  
$\epsilon=a^+ \epsilon_+ + a^- \epsilon_-$ of the two covariantly 
constant spinors $\epsilon_+$ and $\epsilon_-$ of the 
original  $N=2$ supersymmetry. Let us denote the relative phase 
of $a^+$ and $a^-$ by $a^-/a^+=-ie^{i\theta_B}$ while the 
absolute magnitude can be fixed by the normalization of $\Omega$.
{} From $\int J^3 =\frac{3i}{2}e^{-2U}\int \Omega\wedge\bar\Omega$
one infers 
\beq\label{Omeganorm}
e^{U}=\sqrt{2}\, e^{\frac{1}{2}(K^{\rm K}-\Kcs)}\ ,
\eeq
where K\"ahler potential $K^{\rm K}(t)$
is given in \eqref{Kks} while $\Kcs(q)$ is the restriction of the K\"ahler 
potential \eqref{csmetric} to the real slice $\cM^{\rm cs}_\bbR$.
The existence of $\epsilon$ imposes constraints
on the map $\varphi$. These BPS conditions read
\beq\label{sLagr-cond}
  \varphi^*(\Omega)\ =\ e^{U+i\theta_B} \sqrt{\det\big({\varphi^*(\hat g+ \hat B_2) +  2\pi \alpha' F_2}\big)} 
                        d^3 \lambda\ , \qquad 
  \varphi^*\Jc + i 2\pi \alpha' F_2\ =\ 0\ ,
\eeq 
where $\Jc$ is given in \eqref{def-t}.
 The second condition in \eqref{sLagr-cond} enforces 
$\varphi^*(J)=0$ as well as $\varphi^*\hat B_2 + 2\pi \alpha' F_2 =0$, such that the first equation 
simplifies to 
\beq \label{cal1}
  \varphi^*\R( e^{-i\theta_B}\Omega)\ =\ e^U \sqrt{\det\big(\varphi^*\hat g\big)} d^3 \lambda\ , \qquad 
  \varphi^*\I( e^{-i\theta_B}\Omega)\ =\ 0\ ,
\eeq
where we have used that the volume element on $\cW_3$ is real. The equations \eqref{sLagr-cond} 
and \eqref{cal1} imply that the Euclidean $D2$ branes have to wrap special Lagrangian cycles 
in $Y$, which are calibrated with respect to 
$\R(e^{-U-i\theta_B}\Omega)$. 
On the other hand, recall 
that the orientifold planes are located 
at the fixed points of the anti-holomorphic 
involution $\sigma$ in $Y$ which are
special Lagrangian cycles calibrated 
with respect to $\R(e^{-U-i\theta}\Omega)$
  as was argued in  eqs.\
\eqref{OLagr} and \eqref{calibr-O6}.\footnote{$e^{-U}$ is the normalization factor which was left undetermined in \eqref{calibr-O6}.}
Thus, in order for the D-instantons to 
preserve the same linear combination of the supercharges as the orientifold, we have to 
demand  $\theta_B =\theta$.
 Using this constraint and inserting the calibration conditions 
\eqref{cal1} back into \eqref{instact} one finds
\beq \label{instact2}
  S_{D2} = -2\mu_3 \, 
  \int_{\cW_3} \varphi^*\big[\R( C\Omega) \big] + i\mu_3 \int_{\cW_3} \varphi^*(\hat C_3)\ = \ 
  -i\, \int_{\cW_3} \varphi^*\Omegac \ ,
\eeq 
where $C=\frac{1}{2} e^{-\phi-i\theta} e^{-U}$ 
was defined in eqs.\ \eqref{def-C}, \eqref{4d-dilaton} and 
$\Omegac$ is given in \eqref{N=1coords}. The coefficients of $\Omegac$ 
expanded in a basis of $H^{3}_+(Y)$
are exactly the $N=1$ K\"ahler coordinates $(N^k,T_\lambda)$ introduced in \eqref{Oexp}. As a consequence the instanton action 
\eqref{instact2} is linear and thus holomorphic in these coordinates
which shows that $D2$-instantons 
can correct the superpotential.
Explicitly such corrections can be obtained by evaluating 
appropriate fermionic 2-point functions which are weighted 
by $e^{-S_{D2}}$ \cite{HM}. Applying \eqref{instact2}
and keeping only the lowest term in the fluctuations 
of the instanton one obtains corrections of the form 
\beq
   W_{D3} \propto  e^{i\int_{\Sigma_3}  \Omegac}\ , 
\eeq
where $\Sigma_3$ is the three-cycle wrapped by the $D2$ instanton. 
This result can be lifted to M-theory by embedding Calabi-Yau orientifolds into 
compactifications on special $G_2$ manifolds.
In this case the $D2$ instantons correspond 
to membranes wrapping three-cycles in the $G_2$ space 
which do not extend in the 
dilaton direction \cite{HM,KMcG}. The embedding of IIA 
orientifolds into $G_2$ manifolds and the comparison of the
respective effective actions is the subject of the next section.


\section{The $G_2$ embedding of IIA orientifolds}\label{G2}

In this section we discuss the relationship between the type IIA 
Calabi-Yau orientifolds considered so far 
and $G_2$ compactifications of  M-theory. 
In refs.\ \cite{KMcG} it was argued that for a specific class
of $G_2$ compactifications $X$, type IIA orientifolds appear at special
loci in their moduli space. 
More precisely,  
these  $G_2$ manifolds have to be such that they admit the form 
\beq \label{spG_2}
  X\ =\ (Y \times S^1)/{\hat \sigma}\ ,
\eeq
where $Y$ is a Calabi-Yau threefold and $\hat \sigma = (\sigma,-1)$ 
is an involution which inverts the coordinates of the circle $S^1$ 
and acts as an anti-holomorphic isometric involution on $Y$. 
$\sigma$ and $\hat \sigma$ can have a non-trivial fixpoint set
and as a consequence $X$ is a singular $G_2$ manifold. 
In terms of the 
type IIA orientifolds the fixpoints of $\sigma$ are the locations 
of the $O6$ planes in $Y$ and as we already discussed earlier cancellation of 
the appearing  tadpoles require the presence 
of appropriate $D6$-branes. In this paper we froze all
excitation of the $D6$-branes and only discussed the effective action
of the orientifold bulk. In terms of $G_2$ compactification this 
corresponds to the limit where $X$ is smoothed out and all additional 
moduli arising in this process are frozen.

The purpose of this section is to check the embedding of type IIA
orientifolds into $G_2$ compactifications of M-theory 
at the level of the $N=1$ effective action.
For orientifolds the effective action was derived in sections
3 and 4 and so as a first step we need to 
recall the effective action of M-theory 
(or rather eleven-dimensional supergravity) on smooth $G_2$ manifolds
\cite{PT, HM, Hitchin1, GPap,BW}. 

The only multiplet in eleven-dimensional 
supergravity is the supergravity multiplet, which consists of the metric $g_{11}$
and a three-form $C_3$ as bosonic components. 
The effective action for these fields is given by \cite{CJS}
\beq \label{11dact}
  S^{(11)} = \frac{1}{ \kappa_{11}^2} \int \tfrac{1}{2} R *\mathbf{1} - \tfrac{1}{4} G_4 \wedge * G_4 
                      - \tfrac{1}{12} C_3 \wedge G_4 \wedge G_4 \ ,
\eeq
where $G_4 = dC_3$ is the field strength of $C_3$. As in the reduction 
on Calabi-Yau manifolds one chooses the background metric 
to admit a block-diagonal form
\beq \label{lin-el}
  ds^2 = ds^2_4(x) + ds^2_{G_2}(y)\ ,
\eeq
where $ds^2_4$ and $ds^2_{G_2}$ are the line elements of a Minkowski
and a $G_2$ metric, respectively.
The Kaluza-Klein Ansatz 
for the three-form $C_3$ reads 
\beq
  C_3 = c^i(x)\, \phi_i + A^\alpha(x) \wedge \omega_\alpha \ , \qquad i=1,\ldots,b^3(X)\ ,\quad \alpha = 1, \ldots, b^2(X)
\eeq
where $c^i$ are real scalars and $A^\alpha$ are one-forms in four space-time dimensions.
The harmonic forms $\phi_i$ and $\omega_\alpha$ span a basis of
$H^3(X)$ and $H^2(X)$, respectively.  
The $G_2$ holonomy allows for exactly one covariantly 
constant spinor which can be used to define a real, harmonic and 
covariantly constant 
three-form $\Phi$.\footnote{The covariantly constant
three-form is the analog of the holomorphic three-form $\Omega$ 
on Calabi-Yau manifolds.}
The deformation space of the $G_2$ metric has dimension $b^3(X)=\dim H^3(X,\bbR)$ and 
can be parameterized by expanding $\Phi$ 
into the basis $\phi_i$ \cite{Joyce} 
\beq
   \Phi = s^i(x)\, \phi_i \ .
\eeq
One combines the real scalars
$s^i$ and $c^i$ into complex coordinates according to
\beq
  S^i = c^i + i s^i\ ,
\eeq
which form the bosonic components of $b^{3}(X)$ chiral multiplets.
In addition the  effective four-dimensional supergravity features
$b^{2}(X)$ vector multiplets with the $A^\alpha$ as bosonic components. 
Due to the $N=1$ supersymmetry, 
the couplings of these multiplet are again expressed in terms of 
a K\"ahler potential $K_{G_2}$,  gauge-kinetic 
coupling functions $f_{G_2}$ and a (flux induced) superpotential $W_{G_2}$.
Let us discuss these functions in turn.

The K\"ahler potential was found to be \cite{HM,Hitchin1,GPap,BW}
\beq \label{G_2Kpo}
  K_{G_2}\ =\ - 3 \ln \big(  \tfrac{1}{ \kappa^2_{11}} \tfrac{1}{7} \int_X \Phi \wedge * \Phi \big)\ ,  
\eeq 
where $\frac{1}{7}\int \Phi \wedge * \Phi = \text{vol}(X)$ is the volume of the $G_2$ manifold $X$. 
The associated K\"ahler metric is given by 
\beq \label{G_2Kmetr}
  \partial_{i}\bar \partial_{\bj}  K_{G_2}\ =\ \tfrac{1}{4} \text{vol}(X)^{-1} \int_X \phi_i \wedge * \phi_j\ ,
  \qquad 
  \partial_{i}  K_{G_2}\ =\ \tfrac{i}{2} \text{vol}(X)^{-1} \int_X \phi_i \wedge * \Phi\ ,
\eeq
and obeys the no-scale type condition
\beq
  (\partial_{i}  K_{G_2})\,  K_{G_2}^{i \jb}\, (\partial_{\bj}  K_{G_2})  = 7\ . 
\eeq 

The holomorphic gauge coupling functions 
$f_{G_2}$ arise from the couplings of $C_3$ in 
\eqref{11dact}. At the tree level they are linear in $S^i$ 
and read \cite{HM,GPap}
\beq \label{gauge-kinG}
  (f_{G_2})_{\alpha \beta} = \tfrac{i}{2 \kappa_{11}^2}\, 
S^i\int_X \phi_i \wedge \omega_\alpha \wedge \omega_\beta\ . 
\eeq

Finally, non-vanishing background flux of $G_4$ 
induces a scalar potential which via \eqref{N=1pot} 
can be expressed in terms of the superpotential 
\cite{Gukov,AS,BW}
\beq \label{G_2supo}
  W_{G_2}\ =\  \tfrac{1}{4 \kappa_{11}^2}\int_X \big(\tfrac{1}{2} C_3 +i\Phi) \wedge G_4\ .
\eeq 
(The factor $1/2$ ensures holomorphicity of $W_{G_2}$ in the
coordinates $S^i$ and compensates the quadratic dependence on $C_3$
\cite{BW}.)

In order to compare the low energy effective theory of $G_2$
 compactifications
with the one of the orientifold we first have to restrict to the special $G_2$ manifolds 
$X$ introduced in \eqref{spG_2}. 
This can be done by analyzing how the cohomologies of $X$ are related to the 
ones of $Y$. As in equation \eqref{cohom-split} we consider the splits $H^p(Y)=H^p_+ \oplus H^p_-$ 
of the cohomologies into eigenspaces
of the involution $\sigma$. Working on the $G_2$ manifold $X$ given in \eqref{spG_2} 
we thus find the $\hat \sigma$-invariant cohomologies
\beq \label{splcoho}
   \begin{array}{cclcrcl}
   H^2(X) &=& H^2_+(Y)\ , &\ &
   H^3(X) &=& H^3_+(Y) \oplus \big[H^{2}_-(Y)\wedge H_-^1(S^1)\big] \ ,\Big. \\
   H^5(X) &=& H^4_-(Y)\wedge H_-^1(S^1)\ , &&
   H^4(X) &=& H^4_+(Y) \oplus \big[H^{3}_-(Y)\wedge H_-^1(S^1)\big] \ ,   
  \end{array}
\eeq
where $H^2(X)$ and $H^5(X)$ as well as $H^3(X)$ and $H^4(X)$ are
Hodge duals. $H_-^1(S^1)$ is the one-dimensional space containing
the odd one-form of $S^1$. The split of $H^3(X)$ induces a split of the $G_2$-form
$\Phi$ which is most easily seen by introducing locally an orthonormal basis
$(e^1,\ldots,e^7) \in \Lambda^1(X)$ of one-forms. 
In terms of this basis one has \cite{Joyce,Hitchin1,CS}
\beq\label{Phidecomp}
  \Phi
      \ =\ J_M \wedge e^7 + \text{Re} \Omega_M \ ,
\qquad
*\Phi= \tfrac12 J_M\wedge J_M + \I \Omega_M\wedge e^7\ , 
\eeq
where 
\bea \label{defJO}
  J_M = e^1\wedge e^2 + e^3\wedge e^4 + e^5\wedge e^6\ , \quad 
\Omega_M = (e^1 + ie^2)\wedge(e^3+ie^4)\wedge(e^5+ie^6)\ .
\eea
Applied to the manifold \eqref{spG_2} 
we may interpret $e^7=dy^7$ as being the odd one-form
along $S^1$. Since  $\Phi$ is required to be invariant under 
$\hat\sigma$
and $\sigma$ is anti-holomorphic the decomposition 
\eqref{Phidecomp} implies
\beq \label{splitPhi}
  \hat \sigma^* J_M = - J_M\ , \qquad  
\hat \sigma^* \Omega_M = \bar \Omega_M\ .
\eeq  
In terms of the 
basis vectors $e^1,\ldots,e^6$ this is ensured by choosing
$e^4,e^5,e^6$ to be odd  and
$e^1,e^2,e^3$ to be even under $\sigma$.
We see that  $J_M$ and $\Omega_M$ satisfy
the exact same conditions as the corresponding forms of the
orientifold
(c.f.\ \eqref{constrJ}, \eqref{constrO}) and thus have to be proportional to
$J$ and $C\Omega$ used in section \ref{IIAorientifolds}. In order to
determine the exact relation  
it is neccesary to fix their relative normalization. 
The relation between
$J_M$ and the  K\"ahler form $J$ in the string frame 
can be determined from the relation of the respective metrics.
Reducing eleven-dimensional supergravity to type IIA supergravity in
the string frame 
requires the line element \eqref{lin-el} of the eleven-dimensional 
metric to take the form 
\beq \label{metransatz}
  ds^2 = e^{-{2 \hat \phi}/{3}} ds_4^2(x) + 
         e^{-{2 \hat \phi}/{3}} g_{(s)\, ab}\, dy^a dy^b + e^{{4 \hat \phi}/{3}} (dy^7)^2\ ,
\eeq   
where $a,b=1,\ldots,6$. 
The factors $e^{\hat \phi}$ of  the ten-dimensional dilaton are
chosen such that the type IIA
supergravity action takes the standard form with 
$g_{(s)}$ being the Calabi-Yau metric in string frame
(see e.g.~\cite{JPbook}). 
Consequently we have to identify 
\beq\label{JM}
J_M=e^{-{2 \hat \phi}/{3}} J\ .
\eeq

Similarly, using \eqref{defJO} we find that the normalization of $\Omega_M$ is given by
\bea \label{norm}
  J_M \wedge J_M \wedge J_M  = \frac{3i}{4}\, \Omega_M \wedge \bar \Omega_M\ .
\eea 
Integrating over $Y$ and using \eqref{JM}, \eqref{Kks} and \eqref{csmetric} we obtain
\bea \label{normO}
  \Omega_M =  e^{-\hat \phi-i\theta} 
  e^{\frac{1}{2}(\Kcs - K^{\rm K})}\, \Omega =  \sqrt{8} C\Omega\ ,
\eea
where $C$ is given in \eqref{def-C}.
The phase $e^{i\theta}$ drops out in \eqref{norm} such 
that we can choose it
as in \eqref{constrO} in order to fulfill \eqref{splitPhi}.
Inserting $J_M$ and $\Omega_M$ into equation \eqref{splitPhi} 
one arrives at
\beq \label{Phi-o}
  \Phi = J \wedge d\tilde y^7 + \sqrt{8} \text{Re}(C\Omega) \ ,
\eeq
where we defined $d\tilde y^7 = e^{-\frac{2 \hat \phi}{3}} dy^7$. The form 
$d\tilde y^7$ is normalized such that $\int_{S^1} d\tilde y^7=2\pi R$ where 
the metric \eqref{metransatz} was used and 
$R$ is the $\phi$-independent radius of the internal circle. 
We also set $\kappa^2_{10}=\kappa_{11}^2 / 2\pi R = 1$ henceforth.
Using \eqref{Phi-o}, \eqref{Phidecomp} 
and \eqref{def-C} we calculate 
\beq \label{voldec}
          \tfrac{1}{\kappa^2_{11}}\, \tfrac{1}{7} \int \Phi \wedge * \Phi 
           = e^{-\frac{4\hat \phi}{3}} \, \tfrac{1}{6} \int J \wedge J \wedge J \ , 
\eeq
which  equivalently  can be obtained by applying the volume split 
$\text{vol}(X)=\text{vol}(Y)\cdot\text{vol}(S^1)$ evaluated in the metric \eqref{metransatz}.
Inserting \eqref{voldec} into \eqref{G_2Kpo} using \eqref{def-C}
we obtain
\beq\label{IIAori}
  K_{G_2}\ =\
       - \ln \Big[  \tfrac{1}{6} \int J \wedge J \wedge J \Big]
       - 2\ln\Big[2\int_Y \R(C \Omega)\wedge *_6 \R(C\Omega)\Big]\ .
\eeq
 Thus we find exactly the K\"ahler potential 
$K$ of the type IIA orientifold as given in \eqref{N=1Kpot}.\footnote{%
In terms of the Hitchin functionals \cite{Hitchin1} recently discussed in 
\cite{DGNV,Nekrasov} the reduction of the 
$G_2$ K\"ahler potential \eqref{G_2Kpo} corresponds to
the split of the seven-dimensional Hitchin functional to the 
two six-dimensional ones \ref{IIAori}.}

In order to compare the gauge kinetic functions and the superpotential
we also need to identify the 
K\"ahler coordinates of the two theories. 
$C_3$ splits under the decomposition \eqref{splcoho} 
of the cohomologies as\footnote{We have introduced a factor of $\sqrt{2}$ for later convenience.} 
\beq \label{spl-C}
  C_3 = \hat B_2 \wedge d\tilde y^7 + \sqrt{2} \hat C_3 \ ,
\eeq
where $\hat B_2$ is an odd two-form on $Y$ 
and $\hat C_3$ an even three-form on $Y$.
Combining \eqref{Phi-o} and \eqref{spl-C} using \eqref{N=1coords}
one finds
\beq \label{spl-CPhi}
  S^i \phi_i\ =\ C_3 + i\Phi\ =\ \Jc \wedge d\tilde y^7 + \sqrt{2}\, \Omegac\ .
\eeq
As discussed after
\eqref{N=1coords} the coefficients arising in the expansions of $\Jc$ and $\Omegac$
into the basis $(\alpha_k,\beta^\lambda)$ of $H^{3}_+(Y)$
and $\omega_a$ of $H^2(Y)$ are exactly the orientifold coordinates and 
therefore we have to identify
$S^a \cong t^a$ and $S^K \cong (N^k,T_\lambda)$.
With this information at hand, it is not difficult to show that the 
gauge-kinetic couplings \eqref{gauge-kinG} coincide with \eqref{gauge-A}. 
One splits $\phi_a = \omega_a \wedge d\tilde y^7$ and obtains 
\beq
  (f_{G_2})_{\alpha \beta} 
= \tfrac{i}{2} S^a \int_Y \omega_a \wedge \omega_\alpha \wedge
\omega_\beta\ \sim i t^a \cK_{a\alpha\beta} = (f_{OY})_{\alpha \beta}\  
 ,
\eeq 
where the precise factor depends on the normalization of
the gauge fields.

It remains to compare the flux induced superpotentials \eqref{G_2supo} 
with \eqref{superpot1}. Using the 
cohomology splits \eqref{splcoho} and \eqref{spl-C} 
the background flux 
splits accordingly as $G_4 = H_3 \wedge d\tilde y^7 + \sqrt{2} F_4$. 
Inserted into \eqref{G_2supo}
using \eqref{spl-CPhi} we arrive at
\beq
  W_{G_2} = \tfrac{1}{\sqrt{8}}\int_Y \Jc \wedge F_4 + 
            \tfrac{1}{\sqrt{8}}\int_Y \Omegac \wedge H_3
\eeq
Compared to \eqref{superpot1} the superpotential $W_{G_2}$ only 
includes terms proportional to the fluxes $H_3$ and $F_4$.\footnote{%
The term proportional to $e_0$ in \eqref{superpot2} can be absorbed
into a redefinition of $\R t^a$ \cite{BW}.} 
The remaining terms in \eqref{superpot1} should arise once manifolds with
$G_2$ structure (instead of $G_2$ holonomy) are considered.
However, the discussion of this generalization is beyond the scope of
this paper.


\section{Mirror symmetry}\label{mirrorsec}

In this section we discuss mirror symmetry 
for Calabi-Yau orientifolds from the point of view of the effective
action derived in the large volume limit. More precisely, we compare the $N=1$
data obtained in the previous sections  for 
type IIA orientifolds with the ones 
determined in ref.\ \cite{GL} for type IIB orientifolds. 
In order to do so we need to briefly review some properties of 
type IIB Calabi-Yau orientifolds \cite{AAHV,BH,GL}. 

Similar to type IIA the type IIB orientifolds are obtained 
by modding out IIB string theory compactified on a 
Calabi-Yau manifold $\tilde Y$ by a discrete 
symmetry ${\cal O}$ which is involutive ${\cal O}^2 = 1$ and 
includes worldsheet parity $\Omega_p$. 
For type IIB one has two distinct choices 
for ${\cal O}$ depending on the transformation 
properties of the Calabi-Yau three-form $\Omega$.
They are given by \cite{AAHV,BH}
\beq\label{constrOB}
\begin{tabular}{lllll}
${\cal O}_1$ &=& $\Omega_p \sigma_B (-)^{F_L}\ ,\qquad$ & 
$\sigma_B^* \Omega\ =\ - \Omega\ ,\qquad$  &O3/O7 ,\\[1ex]
${\cal O}_2$ &=& $\Omega_p \sigma_B$ \ , 
 &$ \sigma_B^* \Omega\ =\ \Omega\ ,$&   O5/O9 .
\end{tabular}
\eeq
Modding out by ${\cal O}_1$ leads to the presence of
 $O3/O7$ planes  while modding out by ${\cal O}_2$ results in  
$O5/O9$ planes.
$\sigma_B$ is again an involutive symmetry $\sigma_B^2=1$
which acts on the Calabi-Yau
coordinates but in contrast to the situation in type IIA it is a holomorphic 
isometry of $\tilde Y$ and therefore obeys in both cases
$\sigma_B^* J = J$.

The $N=1$ spectrum is obtained from the 
invariant modes of the ten-dimensional type IIB fields 
$\phi_B, \hat C_0, \hat B_2, \hat C_2$ and $\hat C_4$.
Without repeating the details
one finds that in analogy to \eqref{fieldtransf} the invariant modes
have to transform according to \cite{BH}
\begin{equation} \label{fieldtransfB}
\begin{array}{lcl}
\\
\sigma^*_B  \hat \phi &=& \  \hat \phi\ , \\
\sigma^*_B   \hat g &=& \ \hat g\ , \\
\sigma^*_B   \hat B_2 &=& -  \hat B_2\ ,
\end{array}
\hspace{1cm}
\begin{array}{lcl}
\multicolumn{3}{c}{ \underline{O3/O7}} \\[2ex]
\sigma^*_B  \hat C_0 &=& \ \ \hat C_0\ , \\
\sigma^*_B   \hat C_2 &=& - \hat C_2\ , \\
\sigma^*_B   \hat C_4 &=& \ \ \hat C_4\ , 
\end{array}
\hspace{1cm}
\begin{array}{lcl}
\multicolumn{3}{c}{ \underline{O5/O9}} \\[2ex]
\sigma^*_B   \hat C_0 &=& - \hat C_0\ , \\
\sigma^*_B   \hat C_2 &=& \ \ \hat C_2\ , \\
\sigma^*_B   \hat C_4 &=& - \hat C_4\ , 
\end{array}
\end{equation}
where the first column is identical for both involutions $\sigma_B$ 
in \eqref{constrOB}. 

Since $\sigma_B$ is a holomorphic involution 
the cohomologies of $\tilde Y$ split again into eigenspaces of $\sigma_B$
as
\beq \label{cohom-splitB}
  H^{(p,q)} = H^{(p,q)}_+ \oplus  H^{(p,q)}_-\ .
\eeq
In the Kaluza-Klein reduction on $\tilde Y$, the ten-dimensional fields
are expanded in harmonic forms in the appropriate eigenspaces of $\sigma_B$.
Inserting these expansions into the ten-dimensional 
IIB supergravity action results in an 
$N=1$ supergravity in $d=4$ which can be 
brought into the form \eqref{N=1action}
and therefore is characterized by a K\"ahler potential, a set of gauge-kinetic
functions and a superpotential.  For both cases ($O3/O7$ and $O5/O9$)
these $N=1$ data have
been determined in ref.\ \cite{GL} and we recall the results as we go along.

Analogously to \eqref{direct-mod} the moduli space of
type IIB orientifolds  locally is a direct product
of two K\"ahler manifolds  
\beq \label{modsp-B}
  \tilde \cM^{\rm K}_B \times \tilde \cM^{\rm Q}_B\ ,
\eeq 
where $\tilde \cM^{\rm K}_B$ is again a special K\"ahler manifold obtained 
by reducing the type IIB $N=2$ special K\"ahler manifold while
$\tilde \cM^{\rm Q}_B$ is a K\"ahler subspace of the $N=2$ quaternionic
manifold. However, in type IIB
the manifold $\tilde \cM^{\rm K}_B$ is spanned by the complex 
structure deformations of $\tilde Y$ respecting the constraints
\eqref{constrOB}. This implies that it can be parameterized by 
$h^{(2,1)}_-$ complex scalars $z^a$ for orientifolds with $O3/O7$ planes 
and $h^{(2,1)}_+$ complex scalars $z^\alpha$ for orientifolds with $O5/O9$ planes.
$\tilde \cM^{\rm Q}_B$ has complex dimension $h^{(1,1)}+1$ 
for both type IIB theories and includes the type IIB dilaton,
the K\"ahler deformation of $\tilde Y$ and 
the scalars arising from $\hat B_2, \hat C_2$ and $\hat C_4$. 
Additionally the IIB effective theory contains 
$h^{(2,1)}_+ (h^{(2,1)}_-)$ vector multiplets for orientifolds 
with $O3/O7 (O5/O9)$ planes.  We summarize
the number of chiral and vector multiplets in table \ref{numberM}.\\
\begin{table}[h] 
\begin{center}
\begin{tabular}{|l|c|c|c|} \hline 
 \rule[-0.3cm]{0cm}{0.8cm} 
multiplets& IIA$_Y$ \  $O6$ & IIB$_{\tilde Y}$ \  $O3/O7$ & IIB$_{\tilde Y}$ \  $O5/O9$ \\ \hline\hline
 \rule[-0.3cm]{0cm}{0.9cm} 
 {vector multiplets} &   $h_+^{(1,1)}$ & $h_+^{(2,1)}$ & $h_-^{(2,1)}$ \\ \hline
\rule[-0.3cm]{0cm}{0.9cm} 
 chiral multiplets in $\tilde \cM^{\rm K}$& $h_-^{(1,1)}$ & $h_-^{(2,1)}$ & 
                      $h_+^{(2,1)} $   \\ \hline
 \rule[-0.3cm]{0cm}{0.9cm} 
 chiral multiplets in $\tilde \cM^{\rm Q}$&$h^{(2,1)} + 1$&$h^{(1,1)} + 1$ & 
                      $h^{(1,1)} + 1$   \\ \hline
\end{tabular} 
\caption{Number of $N=1$ multiplets of orientifold compactifications.}\label{numberM}
\end{center}
\end{table} 

Since we want to discuss mirror symmetry we choose $\tilde Y$ 
to be the mirror manifold of $Y$. This implies that the
non-trivial Hodge numbers $h^{(1,1)}$ and $h^{(2,1)}$ of $Y$ and $\tilde Y$ 
satisfy
\beq\label{mirror}
  h^{(1,1)}(Y)\ =\ h^{(2,1)}(\tilde Y)\ , \qquad h^{(2,1)}(Y)\ = \ h^{(1,1)}(\tilde Y)\ . 
\eeq   
In addition, we also have to specify the involutions $\sigma_A$ and 
$\sigma_B$ which are identified under mirror symmetry. Since the
discussion
in this paper is quite generic and never specified any involution
$\sigma$ explicitly we also keep the discussion of mirror symmetry
generic. That is we assume that there exists a mirror pair
of manifolds $Y$ and $\tilde Y$ with a mirror pair of involutions
$\sigma_A, \sigma_B$. This implies an orientifold version of
\eqref{mirror},\footnote{For the sector of $\tilde \cM^{\rm Q}$ mirror
  symmetry
is a constraint on the couplings rather than the Hodge
numbers.}  i.e.\
\bea \label{matchchohm}
   O3/O7&: &\quad h^{1,1}_-(Y) = h^{2,1}_-(\tilde Y) \ , \qquad  h^{1,1}_+(Y) = h^{2,1}_+(\tilde Y) \ ,\nn\\
   O5/O9&: &\quad h^{1,1}_-(Y) = h^{2,1}_+(\tilde Y)   \ , \qquad h^{1,1}_+(Y) = h^{2,1}_-(\tilde Y) \ .
\eea

Our next task will be to match the couplings of the mirror theories.
Since the effective actions on both sides 
 are only computed in the large volume limit
we can expect to find agreement only if we also take 
the large complex structure limit exactly as in the $N=2$ mirror
symmetry.
However, if one believes in mirror symmetry one can use the 
the geometrical results of the complex structure moduli space to
`predict' the corrections to its mirror symmetric component.
This is not quite as straightforward since the full $N=1$ moduli space is a
lot more complicated than the underlying $N=2$ space \cite{BH}.
Let us therefore start our analysis with the simpler situation of the 
special K\"ahler sectors $\tilde \cM^{\rm K}_A,\, \tilde \cM_B^{\rm K}$ and 
the vector multiplet couplings 
and postpone the analysis of $\tilde M^{\rm Q}_{A,B}$ 
to sections \ref{O3O7mirror} and \ref{O5O9mirror}.   

\subsection{Mirror symmetry in $\tilde \cM^{\rm K}$}\label{mirrorMK}
Recall that the manifold $\tilde \cM^{\rm K}_A$ is spanned by the 
complexified K\"ahler deformations $t^a$ preserving the constraint 
\eqref{constrJ}.  Under mirror symmetry these moduli are mapped 
to the complex structure deformations which respect the constraint
\eqref{constrOB}.
In both cases the K\"ahler potential is merely a truncated
version of the $N=2$ K\"ahler potential and one has
\beq
  K^{\rm K}_A \ =\ - \ln \Big[\tfrac{4}{3}  \int_Y J \wedge J \wedge J\Big]
  \quad  \leftrightarrow \quad 
  K^{\rm cs}_B\ =\ -  \ln \Big[-i \int \Omega \wedge \bar \Omega \Big]\ .
\eeq
Both K\"ahler potentials can be expressed in terms of prepotentials
$f_A(t), f_B(z)$ and in the large complex structure limit
$f_B(z)$ becomes cubic and agrees with $f_A(t)$.
Mirror symmetry therefore equates these prepotentials 
and exchanges $J^3$ with $\Omega\wedge\bar\Omega$
exactly as in $N=2$ 
\beq\label{mirrorK}
f_A(t) =  f_B(z) \ , \qquad J^3 \leftrightarrow
\Omega\wedge\bar\Omega\ .
\eeq
Thus for $\tilde \cM^{\rm K}$ mirror symmetry is a truncated
version of $N=2$ mirror symmetry. As we will see momentarily this also
holds for the couplings (the gauge kinetic couplings and the
superpotential) which depend on the moduli spanning $\tilde \cM^{\rm K}$.

In type IIA the gauge-kinetic couplings
are given in \eqref{gauge-A} and read
$f_{\alpha \beta}(t) = i\cK_{\alpha\beta c}  t^c$.
The IIB couplings were determined in ref.\ \cite{GL}  to be 
\bea \label{gauge-B}
 f_{\alpha\beta}(z^a) = - {i} \bar \cM_{\alpha\beta}
   = - i\cF_{\alpha\beta}
\ ,
\eea
where in order to not overload the notation we are using the same indices
for both cases.\footnote{We rescaled the type IIB gauge bosons
by $\sqrt 2$ in order to properly match the normalizations.} 
More precisely we are choosing
\bea
\alpha, \beta = 1, \ldots, h^{(2,1)}_+(\tilde Y)\ ,\qquad 
a, b = 1, \ldots, h^{(2,1)}_-(\tilde Y)\ , \qquad \textrm{for} \quad O3/O7\ ,\nn\\ 
\alpha, \beta = 1, \ldots, h^{(2,1)}_-(\tilde Y)\ ,\qquad 
a, b = 1, \ldots, h^{(2,1)}_+(\tilde Y)\ ,\qquad \textrm{for} \quad
O5/O9\ . 
\eea
The matrix $\cF_{\alpha\beta}(z^a)$ is  
holomorphic and the second derivatives of the prepotential restricted
to $\tilde \cM^{\rm  K}_B$. In the large complex structure limit 
$\cF_{\alpha\beta}$ is linear
in $z^a$ and therefore also agrees with the type IIA mirror
couplings. 
Thus mirror symmetry implies the map
\beq 
 \cN_{\alpha \beta}(\bar t^a) \ = \  \cM_{\alpha\beta}(\bar z^a) \ ,
\eeq
in both cases.

It is also straightforward to match the superpotentials which are
induced by RR background flux. For both type IIB cases they are
given by \cite{GVW}
\beq\label{WIIB}  
W_B(z^a) = \int_{\tilde Y} \Omega \wedge F_3\ ,
\eeq
where $F_3$ is the flux of the field strength of $\hat C_2$.
The two-form $\hat C_2$ transforms differently in the two IIB orientifolds
as can be seen in \eqref{fieldtransfB}. Therefore $F_3$
sits in $H^{3}_-(\tilde Y)$ and is determined in terms of
 $2h^{(2,1)}_-+2$ real flux parameters 
for the $O3/O7$ case and sits in $H^{3}_+(\tilde Y)$ 
depending on $2h^{(2,1)}_+ + 2$ real flux parameters
for the $O5/O9$ case. 
On the IIA side the superpotential $W^K(t^a)$ is given in 
\eqref{superpot2} and can be succinctly written as \cite{GVW,Gukov}
\beq
W_A(t) = \int e^{\Jc} \wedge F_{RR}\ ,
\eeq
where $F_{RR}$ stands for a formal sum over all even 
RR-fluxes. It 
depends on $2h^{(1,1)}_- + 2$ RR fluxes $(e_\ah,m^\ah)$ in agreement
with \eqref{matchchohm}.
Furthermore, the functional dependence of the 
superpotentials  coincide
under the mirror map \eqref{mirrorK} which more generally can also 
be written as \cite{FMM}
\beq\label{miracle}
e^{\Jc}(t) \leftrightarrow \Omega(z) \ , \qquad F_{RR} 
\leftrightarrow F_3\ .
\eeq

This concludes our discussions of mirror symmetry
for the chiral multiplets which span $\tilde\cM^{\rm K}$.
We have shown that 
the K\"ahler potential, 
the gauge-kinetic coupling functions and the RR superpotential 
agree in the large complex structure limit under mirror symmetry.
In this sector the geometrical quantities on the type IIB side include
corrections which are believed to 
compute worldsheet non-perturbative effects  
such as worldsheet instantons on the type
IIA side. 
This is analogous to the situation
in $N=2$ and  may be traced back to the 
fact, that it is still possible to formulate a topological 
A model counting 
world-sheet instantons for Calabi-Yau orientifolds \cite{AAHV,BFM}.

\subsection{Mirror symmetry in $\tilde \cM^{\rm Q}$}
Let us now turn to the discussion of the K\"ahler manifolds $\tilde \cM^{\rm Q}_{A}$ and
$\tilde \cM^{\rm Q}_{B}$ arising in 
the reduction of the quaternionic spaces. 
On the IIA side the K\"ahler potential is given in \eqref{N=1Kpot}
which is expressed in terms of the $h^{(2,1)}+1$ coordinates
$(N^k,T_\lambda)$ defined in \eqref{Oexp}.
In this definition we did not fix the scale invariance \eqref{real_K}
$\Omega\to
\Omega e^{-\R (h)}$ or in other
words we defined the coordinates in terms of the scale invariant
combination $C\Omega$. Somewhat surprisingly there seem to be two 
physically inequivalent ways to fix this scale invariance.
In $N=2$ one uses the scale invariance to define special 
coordinates $z^K = Z^K/Z^0, z^0 = 1$ where  $Z^0$ is the coefficent
in front of the base element $\alpha_0$. The choice of $Z^0$
is convention and 
due to the symplectic invariance any other choice would be
equally good. 
However, as we already discussed in section 3.1 and 3.3 the 
constraint \eqref{constrO} breaks the symplectic invariance and  
$H^3$ decomposes into two eigenspaces $H^3_+\oplus H^3_-$.
Thus in \eqref{decompO2} we have the choice to scale one of the $Z^k$ 
equal to one or 
one of the $Z^\lambda$ equal to $i$.
Denoting the corresponding basis element by $\alpha_0$, 
these two choices are characterized by 
$\alpha_0 \in H^{3}_+$ or $\alpha_0 \in H^{3}_-$.
This choice identifies the dilaton direction inside the moduli space
and therefore is crucial in identifying the type IIB
mirror. This is related to the fact that in type IIB
the dilaton reside in a chiral multiplet for $O3/O7$ orientifolds and in a
linear multiplet for $O5/O9$ orientifolds.  
Let us discuss these two cases in turn.

%
%

\subsubsection{The Mirror of IIB orientifolds with $O3/O7$ planes}
\label{O3O7mirror}

For type IIB Calabi-Yau orientifolds with 
$O3/O7$ planes the low energy theory was derived in ref.\
\cite{GL}. 
The K\"ahler manifold $\tilde \cM^{\rm Q}_B$ is spanned by 
$h^{(1,1)}(\tilde Y) + 1$ chiral multiplets which arise 
from the expansion of $J, \hat B_2, \hat C_2$ and $\hat C_4$ 
\bea\label{expansionB}
\hat B_2 &=& b^k(x)\, \omega_k\ ,\qquad \hat C_2\ =\ c^k(x)\, \omega_k\ , 
\quad k=1,\ldots, h_-^{(1,1)}(\tilde Y)\ , \\
  J &=&  v^{\lambda}(x)\, \omega_{\lambda}\ ,\qquad   \hat C_4 =  
 \rho_\lambda(x)\ \tilde \omega^\lambda\ ,\quad \lambda = 1,\ldots,h_+^{(1,1)}(\tilde Y)\ ,\nonumber
\eea
where we only displayed the scalar fields in the expansion.
The proper K\"ahler coordinates were identified as\footnote{%
We have sligthly changed the conventions with respect to \cite{GL}, since the
scalars $v^\alpha$ are now given in string frame.}
\bea\label{Kahlerco}
  \tau &=& C_0 + i e^{-\phi_B}\ , \qquad  G^k\ =\ c^k - \tau
  b^k\ ,
\nn\\
 T_\lambda &=& 2i\rho_\lambda 
+ e^{-\phi_B} \cK_{\lambda \rho \sigma} v^\rho v^\sigma 
- i \cK_{\lambda kl} b^k G^l\ ,
\eea
where $C_0$ is the RR scalar 
and $e^{\phi_B}$ is the type IIB dilaton. 
The intersection numbers $\cK_{\lambda \rho \sigma}$
and $\cK_{\lambda kl}$ are defined exactly as in \eqref{int-numbers} 
and are the only non-vanishing intersections of the even cohomologies
in IIB orientifolds. The K\"ahler potential is given by 
\beq \label{Kpo37}
  K^{\rm Q}_B(\tau,G,T)\ =\ -2\ln \Big[e^{-2\phi_B} \int J\wedge J\wedge J \Big]\ =\
- \ln(e^{-4D_B})\ , 
\eeq
where $e^{D_B}$ is the four-dimensional dilaton.
$K^{\rm Q}$ can only be given implicitly as a function
of  $v^\lambda$ and $e^{\phi_B}$ which are determined by
\eqref{Kahlerco} in terms of the variables 
$\tau, T_\lambda$ and $G^k$.

Now we want to show that in the large complex structure limit
$K^Q_A$ given in \eqref{intKQ} coincides with
$K^{\rm Q}_B$ given in \eqref{Kpo37}.
It turns out that in order to do so we need to choose
$\alpha_0 \in H^3_{+}$ and the dual basis element
$\beta^0\in H^3_{-}$.
It is convenient to keep track of this choice and therefore
we mark the $\alpha$'s and $\beta$'s which contain $\alpha_0$
and $\beta^0$ by putting a hat on the corresponding index. 
Thus we work in the basis $(\alpha_\kh,\beta^\lambda)$ 
of $H^3_+$ and $(\alpha_\lambda,\beta^\kh)$
of $H^{3}_-$. Therefore, we rewrite the combination $C\Omega$ as 
\beq
  C\Omega = g_A^{-1}(\textbf{1}\, \alpha_0 + q^k \alpha_k + iq^\lambda \alpha_\lambda) + \ldots\ ,
\eeq
where we introduced $g_A$ and the real special coordinates 
\beq \label{realspC1}
  g_A =\frac{1}{\R(CZ^0)}\ ,\qquad q^k = \frac{\R(CZ^k)}{\R(CZ^0)}\ , \qquad q^\lambda = \frac{\I(CZ^\lambda)}{\R(CZ^0)}\ .
\eeq
We also need to express the prepotential $\cF(Z)$ 
in the special coordinates $q^k,q^\lambda$.
In anology to \eqref{def-f} one defines a function $f(q)$ 
 such that 
\beq \label{def-h(q)}
  \cF\big(\R[CZ^\kh],i\I[CZ^\lambda] \big)\ =\ i\big(\R[ CZ^0]\big)^2\  f(q^k,q^\lambda) \ . 
\eeq
We are now in the position to rewrite the $N=1$ coordinates 
$N^\kh,T_\lambda$ given in 
\eqref{def-NT} in terms of $g_A$ and the special coordinates $q^K$. 
Inserting \eqref{realspC1} 
into \eqref{def-NT} one obtains
\beq \label{c-in-q37}
  N^0\ =\ \tfrac{1}{2} \xi^0 + i g_A^{-1}\ , \qquad
  N^k\ =\ \tfrac{1}{2} \xi^k + i g_A^{-1} q^k \ , \qquad
  T_\lambda\ =\ i \tilde \xi_\lambda - 2 g_A^{-1} f_\lambda(q)\ ,
\eeq
where $f_\lambda$ is the first derivative of $f(q)$ with respect to $q^\lambda$. 

The final step is to specify $f(q)$ in the large complex structure
limit. 
In this limit the $N=2$ prepotential is known to be
\beq \label{N=2pre}
 \cF(Z) = \tfrac{1}{6} (Z^0)^{-1}{\kappa_{KLM} Z^K Z^L Z^M}\ .
\eeq
Inserted into the orientifold constraints
\eqref{Z=0gen} one infers
\beq \label{vankappa37}
  \kappa_{klm} = \kappa_{\kappa \lambda l} = 0 \ , 
\eeq 
while $\kappa_{\kappa \lambda \mu}$ and $\kappa_{\kappa l m}$ can be non-zero.
Using \eqref{vankappa37}, \eqref{def-h(q)} and \eqref{realspC1} 
we arrive at
\beq\label{fori37}
  f(q)\ =\ - \tfrac{1}{6} \kappa_{\kappa \lambda \mu} q^\kappa q^\lambda q^\rho 
           + \tfrac{1}{2} \kappa_{\kappa kl} q^\kappa q^k q^l\ .
\eeq

In order to continue 
we also have to specify the range the indices $k$ and $\lambda$ 
take on the IIA side.
A priori it is not fixed and can be changed by a symplectic transformation.
Mirror symmetry demands 
\beq \label{na-nb}
 k = 1,\ldots, h^{(1,1)}_-(\tilde Y)\ , \qquad  
\lambda = 1,\ldots,h^{(1,1)}_+(\tilde Y)\ ,
\eeq
or in other words there have to be $h^{(1,1)}_-(\tilde Y)$ 
basis elements $\alpha_k$ and $h^{(1,1)}_+(\tilde Y)$ basis elements
$\beta^\lambda$ in $H^3_+(Y)$. In addition the 
non-vanishing couplings $\kappa_{\kappa \lambda \mu}$ and 
$\kappa_{\kappa l m}$
have to be identified with 
$\cK_{\kappa \lambda \mu}$ and $\cK_{\kappa l m}$ appearing
in the definition of the type IIB chiral coordinates \eqref{Kahlerco}.
With these conditions fullfilled
we can insert \eqref{fori37} into \eqref{c-in-q37} and compare with
\eqref{Kahlerco}. This leads to the identification
\beq
 N^{\kh} = (\tau, G^k) \qquad \textrm{and}\qquad
 T_{\lambda}^A = T_{\lambda}^B\ ,
\eeq
which in terms of the Kaluza-Klein variables corresponds to 
\bea\label{phi=g}
 e^{\phi_B}&=& g_A \ ,\qquad  q^\lambda\ =\ v^\lambda\ ,\qquad  q^k\ =\ -b^k\ ,\nn\\
  \xi_0 &=& 2 C_0\ , \quad \xi^k=2(c^k-C_0 b^k)\ , \\
  \tilde \xi_\lambda &=& 2 \rho_\lambda - \cK_{\lambda kl}c^k b^l +
C_0 \cK_{\lambda kl}b^k b^l\ .\nn
\eea
With these identifications one immediately shows 
$e^{D_A} = e^{D_B}$, where $e^{D_A}$ and
$e^{D_B}$ are the four-dimesional dilatons of the type IIA and IIB theory.
This implies that the K\"ahler potentials \eqref{intKQ} and \eqref{Kpo37} of the two theories coincide in the large volume -- 
large complex structure limit. However, the corrections
away from this limit cannot be properly understood 
from a pure supergravity analysis. It is clear that 
$K^{\rm Q}_A$ includes corrections of the mirror IIB
theory but the precise nature of these corrections remains to be understood.

%
%

\subsubsection{The Mirror of IIB orientifolds with $O5/O9$ planes}
\label{O5O9mirror}

In this section we check mirror symmetry for type IIB orientifolds with 
$O5/O9$ planes. As in the previous section
 we first need to briefly recall the results of
ref.\ \cite{GL}. In this case the Kaluza-Klein 
expansion of the ten-dimensional type IIB
fields change as a consequence of the different transformation
properties given in \eqref{fieldtransfB} and \eqref{expansionB}
is replaced by
\bea\label{expansionB59}
 J &=&  v^{k}(x)\, \omega_{k}\ ,\qquad \hat C_2\ =\ C_2(x) + c^k(x)\, \omega_k\ , 
\quad k=1,\ldots, h_+^{(1,1)}(\tilde Y)\ , \\
\hat B_2 &=& b^\lambda(x)\, \omega_\lambda\ ,\qquad    \hat C_4 =  
 \rho_\lambda(x)\ \tilde \omega^\lambda\ ,\quad \lambda = 1,\ldots,h_-^{(1,1)}(\tilde Y)\ .\nonumber
\eea
The proper K\"ahler coordinates 
which span $\tilde \cM^{\rm Q}$ are the 
$h^{(1,1)}+1$ chiral fields 
\bea\label{def-tA}
  t^k& =& -i e^{-\phi_B} v^k + c^k\ , \nn\\
  A_\lambda &=& 2i \cK_{\lambda\rho k} b^\rho\, t^k  + 2i\rho_\lambda\ , \\
  S &=& \tfrac{1}{3} e^{-\phi_B} \cK + 2ih 
- \tfrac{1}{2}b^\lambda A_\lambda\ ,\nn
\eea
where $h$ is a scalar dual to the four-dimensional two-form $C_2$
defined in \eqref{expansionB59} and 
$\cK=\cK_{\lambda\kappa\rho} v^\lambda v^\kappa v^\rho$.
The K\"ahler potential has the exact same form as for the $O3/O7$ case
and is again given by \eqref{Kpo37} but this time it depends implicitly on the variables 
$S,t^k,A_\lambda$ defined in \eqref{def-tA}.
 
In order to find the same chiral data on the IIA side, we have to examine the 
case where $\alpha_0 \in H^3_{-}$. Therefore we choose a basis 
$(\alpha_k,\beta^{\hat \lambda})$ of $H^3_+$ and $(\alpha_{\hat \lambda},\beta^k)$
of $H^{3}_-$. We rewrite the combination $C\Omega$ in this basis as 
\beq
  C\Omega = g_A^{-1}(i\, \alpha_0 + i q^\lambda \alpha_\lambda  + q^k \alpha_k) + \ldots
\eeq
where we introduced the real special coordinates 
\beq \label{realspC2}
  g_A =\frac{1}{\I(CZ^0)}\ ,\qquad q^k = \frac{\R(CZ^k)}{\I(CZ^0)}\ , \qquad q^\lambda = \frac{\I(CZ^\lambda)}{\I(CZ^0)}\ .
\eeq
Let us also express the prepotential $\cF(Z)$ in terms of $q^k,q^\lambda$. As in $N=2$ one defines a 
function $f(q)$ such that 
\beq \label{def-h59}
  \cF\big(\R[CZ^k],i\I[CZ^{\hat\lambda}] \big) =- i\big(\I[ CZ^0]\big)^2\,  f(q^k,q^\lambda) \ . 
\eeq
We can now rewrite the $N=1$ coordinates $T_{\hat\lambda}, N^k$ 
given in \eqref{def-NT} in terms of 
$q^k,q^\lambda$ and $g_A$ as
\bea \label{c-in-q59}
  N^k &=& \tfrac{1}{2} \xi^k + i g^{-1}_A q^k \ , \qquad T_\lambda = i \tilde \xi_\lambda +2 g^{-1}_A f_\lambda(q)\ , \nn\\
  T_0 &=& i \tilde \xi_0 + 2 g^{-1}_A (2f(q)- f_\lambda q^\lambda - f_k q^k)\ , 
\eea
where $f_\lambda,f_k$ are the first derivatives of $f(q)$ with respect to $q^\lambda$ and $q^k$. 

Going to the large complex structure limit, the $N=2$ prepotential takes the form 
\eqref{N=2pre}. We split the indices as $K=(k,\hat \lambda)$ and apply the constraints 
\eqref{Z=0gen} to find that
\beq \label{vankappa59}
  \kappa_{\kappa \lambda \mu} = \kappa_{\kappa k l} = 0 \qquad \kappa_{klm} \neq 0\ ,\qquad 
  \kappa_{\kappa \lambda l} \neq 0\ .
\eeq 
Using \eqref{vankappa59} and \eqref{def-h59} we can calculate $f(q)$ as 
\beq
  f(q) = \tfrac{1}{6} \kappa_{ k l m} q^k q^l q^m - \tfrac{1}{2} \kappa_{\kappa \lambda k} q^\kappa q^\lambda q^k\ .
\eeq
In order to match the chiral coordinates $T_0,T_\lambda,N^k$ 
with the type IIB coordinates 
of \eqref{def-tA} we need again to specify the range of the indices
on the type IIA side. Obviously we need
\beq \label{na-nb59}
k=1, \ldots, h^{(1,1)}_+(\tilde Y)\ , \qquad  \lambda= 1,\ldots,  h^{(1,1)}_-(\tilde Y)\ ,
\eeq
which is the equivalent of \eqref{na-nb} with the plus and minus sign interchanged. 
Thus the non-vanishing intersections can be identfied with 
$\cK_{klm}$ and $\cK_{\kappa\lambda k}$ on the IIB side.
Inserting $f(q)$ back into the equations \eqref{c-in-q59} for the chiral 
coordinates $N^k,T_{\hat \lambda}$ and demanding \eqref{na-nb59} one can 
compare these to the type IIB coordinates \eqref{def-tA}. 
One identifies 
\beq
 T_{\hat \lambda} = (S,A_\lambda)\ ,\qquad  N^{k} = t^k \ .
 \eeq
In terms of the Kaluza-Klein modes this amounts to the identification
\bea
 g_A &=& e^{\phi_B}\ , \qquad q^k = -v^k\ , \qquad 
q^\lambda = b^\lambda\ ,\qquad 
  \xi^k = 2 c^k\ , \nn \\ 
  \tilde \xi_\lambda &=& 2 \cK_{\lambda \kappa l} c^l b^\kappa + 2\rho_\lambda\ , \qquad 
  \tilde \xi_0 = 2h - \cK_{l\lambda \kappa} c^l b^\lambda b^\kappa - \rho_\lambda b^\lambda\ .
\eea  
With these identifications one shows again $e^{D_A} = e^{D_B}$ and as
a consequence the  K\"ahler potentials agree
in the large volume -- large complex structure limit.

In summary, we found that it is indeed possible to obtain both type IIB 
setups as mirrors of the type IIA orientifolds discussed in section 
\ref{IIAorientifolds}. In analogy to \eqref{miracle}
we found in the $\tilde \cM^{\rm Q}$ component the mirror relation
\bea\label{miracle2}
O3/O7: && \R (C\Omega)\ \leftrightarrow\ e^{-\phi_B}\, \R\, e^{\Jc}\ , \qquad C_3 \leftrightarrow C_{\rm RR}\wedge e^{-\hat B_2}\ ,
\nn \\
O5/O9: && \R (C\Omega)\ \leftrightarrow\ e^{-\phi_B}\, \I\, e^{\Jc} \ , \qquad C_3 \leftrightarrow C_{\rm RR}\wedge e^{-\hat B_2}\ .
\eea
However, the crucial role of the two definitions of special coordinates 
remains to be understood further.

Using the correspondence \ref{miracle2} we can extend the observation 
of section \ref{fluxes_sec} that the proper chiral coordinates 
`linearize' the corresponding
  D-brane instanton action  also to 
type IIB orientifolds \cite{GL}.
One can define the form
\beq \label{cal-coords}
A_p  =  (C_{\text{RR}}\wedge e^{- \hat B_2})_p 
+ i e^{-\phi_B}\, \textrm{Cal}_p\ ,
\eeq
where the 
instantons are calibrated with respect to the $p$-form
$\textrm{Cal}_p$. $(C_{\text{RR}}\wedge e^{-\hat B_2})_p$ is a $p$-form
constructed out of the
formal sum of the ten-dimensional RR forms present
in the orientifold theory.
Expanding $A_p $ in terms of $H^{(p)}_+(Y)$ 
results in chiral coordinates  which linearize 
the $D(p-1)$ instanton action.
These coordinates can already be discovered 
in the orientifold theory since the D-branes
are constructed such that they preserve the same $N=1$ supersymmetry
as the orientifolds.

\section{Conclusions}\label{Conclusion}

In this paper we calculated the four-dimensional
effective action of type IIA
Calabi-Yau orientifolds in the presence of background fluxes. We
restricted
ourselves to Calabi-Yau spaces admitting an anti-holomorphic involutive
symmetry which preserves $N=1$ supersymmetry.
The string theory is modded out by an involutive symmetry 
which includes this
geometric symmetry and thus imposes constraints on the spectrum 
and the couplings of the theory.

We computed the effective action by a Kaluza-Klein analysis
valid in the large volume limit and determined the
chiral variables, the K\"ahler potential, the gauge kinetic function
and the flux-induced superpotential at the tree level.
We found that the moduli space of the $N=1$ theory inherits 
a product structure $\tilde \cM^K \times \tilde \cM^Q$ from
the underlying $N=2$ theory obtained by ordinary Calabi-Yau
compactification of type IIA. $\tilde \cM^K$ is
a special K\"ahler manifold parameterized by the complexified K\"ahler form
$\Jc$ which 
decends from the $N=2$ vector multiplets. 
The second component $\tilde \cM^Q$ is parameterized
by the periods of the `new'  three-form $\Omegac$ 
($= C_3 + 2i \R C\Omega$)
containing the complex structure deformations of the Calabi-Yau orientifold. 
It is a K\"ahler submanifold inside
the quaternionic manifold of $N=2$ and 
has a geometric structure similar to the one of the moduli
space of supersymmetric Lagrangian submanifolds \cite{Hitchin2}.

 
A superpotential $W$ is induced once background fluxes are turned on
which depends on all geometrical moduli.
It splits into the sum of two terms 
with one term depending on the RR fluxes and the complexified
K\"ahler form $J_c$ while the second term
features the NS fluxes and $\Omegac$.
Both terms are expected to receive non-perturbative corrections
from worldsheet- and D-brane instantons.
We showed that the respective actions are linear in the
chiral coordinates and therefore can result in holomorphic
corrections to $W$.


We further discussed the embedding of type IIA
orientifolds into a specific class of 
$G_2$ compactification of M-theory. Neglecting the 
contributions arising from the singularities of the $G_2$ manifold
we were able show agreement between the low energy effective
actions. In the superpotential we only discovered 
the terms which decend from the M-theory four-form $G_4$ 
but we neglected the possibility of geometrical fluxes.

Finally we showed that in the large volume -- large complex structure limit
one finds mirror symmetric effective actions if one compares
type IIA and type IIB supergravity compactified on mirror manifolds
and in addition chooses a set of `mirror involutions'.
For $\tilde\cM^K$ mirror symmetry amounts to a truncated versions
of $N=2$ mirror symmetry in that it still relates
two holomorphic prepotentials. In this case the corrections computed
by mirror symmetry are precisely analogous to the situation in $N=2$.
For $\tilde\cM^Q$
the situation is more involved since the geometry of the moduli
space changes drastically. Nevertheless we were able to show that
mirror symmetry in the large volume - large complex structure limit.
However, understanding 
the nature of the corrections computed by mirror symmetry
appear to be more involved and certainly deserves further study.

\subsection*{Acknowledgments}

We have greatly benefited from conversations with  I.~Benmachiche, I.~Brunner,
V.~Cort\'es, R.~Grimm, 
S.~Gukov, H.~Jockers,  A.~Klemm and S.~Sch\"afer-Nameki.

This work is supported by the DFG -- The German Science Foundation,
the European RTN Program MRTN-CT-2004-503369  and the
DAAD -- the German Academic Exchange Service.

\vspace{1cm}
{\Large \bf Appendix}
\renewcommand{\theequation}{\Alph{section}.\arabic{equation}}
\appendix

\section{$N=2$ special geometry of the Calabi-Yau 
moduli space\label{specialGeom}}

In this appendix we briefly summarize the $N=2$ special geometry
of the Calabi-Yau 
moduli space. A more detailed discussion can be found,
for example, in refs.\ \cite{CdO,Strominger2,Freed,N=2review,CRTV}.
A special K\"ahler manifold $\cM$ is a Hodge-K\"ahler manifold (with line bundle $\cL$)
of real dimension $2n$ with associated 
holomorphic flat $Sp(2n+2,\mathbb{R})$ vector bundle $\mathcal{H}$ over $\cM$. Furthermore 
there exists a holomorphic section $\Omega(z)$ of $\cL$ such that 
\beq\label{N=2KP}
  K(z,\bar z) = - \ln i \big<\Omega(z) , \bar \Omega(\bar z)  \big>\ , \qquad 
  \big<\Omega, \partial_{z^K} \Omega\big> = 0\ , \qquad K=1,\ldots n\ ,
\eeq
where $K$ is the K\"ahler potential of $\cM$ and 
$\big<\cdot,\cdot \big>$ is the 
symplectic product on the fibers. 
This is precisely what one encounters in the moduli space
of the complex structure deformations of a Calabi-Yau manifold
with $\Omega$ being the holomorphic three-form.
In this case 
one is lead to set $n=h^{(2,1)}$ and identify 
the fibers of the associated 
$Sp$-bundle with $H^3(Y,\mathbb{C})$. The symplectic product is given by 
the intersections on $H^3(Y,\bbC)$ as
\beq \label{sympl-f}
  \big<\alpha, \beta \big> = \int_Y \alpha \wedge \beta\ .
\eeq
The K\"ahler covariant derivatives of $\Omega$ are denoted by 
$\chi_K$ as explicitly given in \eqref{Kod-form}.
In terms of the symplectic basis $(\alpha_\Kh, \beta^\Kh)$
introduced in \eqref{symplectic} both
 $\Omega$ and $\chi_K$ enjoy the expansion
\beq \label{def-chi}
  \Omega = Z^\Kh\, \alpha_\Kh - \cF_\Kh\, \beta^\Kh\ , \qquad  
  \chi_K = \chi^\Lh_{K}\, \alpha_\Lh - \chi_{\Lh|K}\, \beta^\Lh\ .
\eeq 
The holomorphic functions $Z^\Kh(z)$ and $\cF_\Kh(z)$
are called the periods of $\Omega$, while $\chi^\Lh_{K}(z,\bar z)$ and $\chi_{\Lh|K}(z,\bar z)$ 
are the periods of $\chi_K$.  In terms of $Z^\Kh,\cF_\Kh$ the K\"ahler potential 
\eqref{N=2KP} can be rewritten as in \eqref{csmetric}.

For every special K\"ahler manifold there exists
a complex matrix $\cM_{\Kh \Lh}(z,\bar z)$ defined as 
\beq \label{def-M}
   \cM_{\Kh \Lh} = (\bar \chi_{\Kh|\bar M}\ \ \cF_\Kh)  (\bar \chi^\Lh_{\bar M}\ \ Z^\Lh)^{-1}\ ,
\eeq
where $\chi^{\Lh}_{K}$ and $\chi_{\Lh| K}$ are given in \eqref{def-chi}.
Furthermore, one extracts from \eqref{def-M} the  
identities
\bea \label{ML-hf}
  \cF_{\Kh} = \cM_{\Kh \Lh} Z^\Lh\ , \qquad   \chi_{\Lh| K} = \bar \cM_{\Lh \Mh} \chi^{\Mh}_{K}\ ,
\eea
which can be used to rewrite \eqref{N=2KP} as 
\bea \label{spconst}
   G_{M \bar N} & = & -2 e^{K} \chi^\Kh_M\, \I \cM_{\Kh \Lh}\, \bar \chi^\Lh_{\bar N}\ , \qquad
   1 \ = \ -2 e^{K} Z^\Kh\, \text{Im}\, \cM_{\Kh \Lh}\,\bar Z^\Lh \ , \\
   0 & = & -2 \bar \chi^\Kh_{\bar M}\, \text{Im}\, \cM_{\Kh \Lh}\,\bar Z^\Lh \ .\nn
\eea

If one assumes that  
the Jacobian matrix $\partial_{z^L}\big(Z^K/Z^0 \big)$ is invertible 
$\cF_\Kh$ is the derivative of a holomorphic prepotential $\cF$ with respect to the periods $Z^\Kh$. 
It is homogeneous of degree two and obeys
\beq
   \cF = \tfrac{1}{2}Z^\Kh \cF_{\Kh}\ , \qquad  \cF_\Kh =\partial_{Z^\Kh} \cF\ , \qquad  
   \cF_{\Kh \Lh} =\partial_{Z^\Kh} \cF_{\Lh}\ ,
\qquad \cF_{\Lh}= Z^\Kh \cF_{\Kh \Lh}\ ,
\eeq
which implies that $\cF_{\Kh \Lh}(Z)$ is invariant
under rescalings of $Z^\Kh$. 
Notice that $\cF$ is only invariant under a restricted 
class of symplectic transformations
and thus depends on the choice of symplectic basis. 

The complex matrix 
$\cM_{\Kh \Lh}$ defined in \eqref{def-M} can be rewritten in terms of the 
periods $Z^\Kh$ and the matrix $\cF_{\Kh\Lh}(Z)$ as
\bea \label{gauge-c}
   \cM_{\Kh \Lh}=\overline{ \mathcal{F}}_{\Kh \Lh}+2i \frac{(\text{Im}\; \mathcal{F})_{\Kh \Mh} Z^\Mh
   (\text{Im}\; \mathcal{F})_{\Lh \Nh}Z^\Nh }{Z^\Nh(\text{Im}\; \mathcal{F})_{\Nh\Mh} 
    Z^\Mh}\ .
\eea

Whenever the 
Jacobian matrix $\partial_{z^L}\big(Z^K/Z^0 \big)$ is invertible 
the $Z^\Kh$ can be viewed as projective coordinates of $\mathbb{P}_{h^{(2,1)}+1}$.
Going to a special gauge, i.e.~fixing the K\"ahler transformations
\eqref{crescale}, one introduces
special coordinates $z^K$ by setting $z^K=Z^K/Z^0$. 
Due to the homogeneity of $\cF$ it is possible to define
a holomorphic prepotential $f(z)$ which only depends on the special
coordinates as
\beq\label{def-f}
  \cF(Z) = (Z^0)^2 f(z)\ .
\eeq
In terms of $f$  the K\"ahler potential  given in \eqref{N=2KP} 
reads
\beq \label{Kinz}
  K\ =\ - \ln i|Z^0|^2 \big[2(f-\bar f)-(\partial_K\, f + \partial_{\bar K} \bar f)(z^K - \bar z^K) \big]\ . 
\eeq

The complexified K\"ahler deformations $t^A$ introduced 
in \eqref{4d-dilaton} are special coordinates of 
a special K\"ahler manifold. 
The  K\"ahler potential of the metric $G_{AB}$ given 
in \eqref{Kmetric} is of the form \eqref{Kinz} with
\beq \label{pre-K}
  f(t)=-\tfrac{1}6 \cK_{ABC} t^A t^B t^C\ .
\eeq
Furthermore, inserting \eqref{pre-K}
into \eqref{gauge-c} using \eqref{def-f} 
one determines the gauge-couplings  $\cN_{\Ah \Bh}(t,\bar t)$ 
to be
\bea \label{def-cN}
  \text{Re} \cN &=& \ \
  \left(\ba{cc}-\frac13 \cK_{ABC}b^A b^B b^C &  \frac12 \cK_{ABC} b^B b^C \\
              \frac12 \cK_{ABC} b^B b^C & - \cK_{ABC}b^C  \ea \right)\ , \nn \\
  \text{Im} \cN &=& -\frac{\cK}{6}
  \left(\ba{cc}1 + 4 G_{AB}b^A b^B & -4 G_{AB}b^B  \\
             - 4 G_{AB}b^B &  4 G_{AB}  \ea \right)\ , \nn \\
  (\text{Im} \cN)^{-1} &=& - \frac{6}{\cK}
  \left(\ba{cc}1 & b^A  \\
         b^A &  \frac14 G^{AB} + b^A b^B \ea \right)\ ,
\eea
where $G_{AB}$ is given in \eqref{Kmetric}.

\section{Supergravity with several linear multiplets} \label{linm}

In this appendix we briefly discuss the dualization of several massless linear
multiplets to chiral multiplets. We only discuss the bosonic component fields and do not include possible couplings to vector 
multiplets. Our aim is to extract the K\"ahler potential 
for the $N=1,d=4$ supergravity theory with all linear multiplets replaced by chiral ones.
Let us begin by recalling the effective action for a set of 
linear multiplets $(L^\lambda, D^\lambda_2)$ couplet to  chiral multiplets
$N^k$. It takes the form\footnote{This action can be obtained by a straight forward 
generalization of the action for one linear multiplet given in \cite{BGG}.}
\bea\label{kinetic}
\cL &=& -\tfrac{1}{2}R*\mathbf{1} - 
  \tilde K_{N^k\bar N^l}\, dN^k \wedge * d \bar N^{l}
  + \tfrac{1}{4} \tilde K_{L^\kappa L^\lambda}\, 
  dL^\kappa \wedge * dL^\lambda \nn\\ 
  && + \tfrac{1}{4} \tilde K_{ L^\kappa L^\lambda}\, dD^\kappa_2 \wedge * dD^\lambda_2
     -  \tfrac{i}2\,  dD^\lambda_2 \wedge 
\big(\tilde K_{L^\lambda N^k}\,dN^k -\tilde K_{l^\lambda \bar N^k}\,d\bar N^k\big)
\ ,
\eea
where  $\tilde K(L,N,\bar N)$ is a function of the scalars $L^\lambda$ and
the chiral multiplets $N^k$. The kinetic potential $\tilde K$ is the 
analog of 
the K\"ahler potential in the sense that it encodes the dynamics of the linear
and chiral multiplets. In order to dualize the linear multiplets $(L^\lambda, D^\lambda_2)$ 
into chiral multiplets $(L^\lambda,\tilde \xi_\lambda)$ one replaces
$dD_2^\lambda$ by the form $D_3^\lambda$ and adds the term 
\beq
\cL \to \cL + \delta \cL\ , \qquad
  \delta \cL\ =\  
  - 2\tilde \xi_\lambda\, dD^\lambda_3\ =\ - 2 D_3^\lambda \wedge d\tilde \xi_\lambda\ ,\ 
\eeq
where $\tilde \xi_\lambda(x)$ is a Lagrange multiplier. Eliminating $\tilde \xi_\lambda$
one finds that $dD^\lambda_3=0$ such that locally $D_3^\lambda=dD_2^\lambda$ as required.
Alternatively one can consistently eliminate $D^\lambda_3$ by inserting 
its equations of motion
\beq
  *D_3^\kappa = 4 \tilde K^{L^\kappa L^\lambda}\Big(d\tilde \xi_\lambda + \tfrac{i}4 
  \big(\tilde K_{L^\lambda N^k}\,dN^k -\tilde K_{L^\lambda \bar N^k}\,d\bar N^k\big)\Big) 
\eeq
back into the Lagrangian \eqref{kinetic}. 
The resulting dual Lagrangian takes the form
\bea \label{eff_act1}
\cL &=& -\tfrac{1}{2}R*\mathbf{1} - 
  \tilde K_{N^k\bar N^l}\, dN^k \wedge * d \bar N^{l}
  + \tfrac{1}{4} \tilde  K_{L^\kappa L^\lambda}\, 
  dL^\kappa \wedge * dL^\lambda  \\ 
  && + 4 \tilde K^{L^\kappa L^\lambda} \Big(d\tilde \xi_\kappa - \tfrac{1}2
  \I \big(\tilde K_{L^\kappa N^l}\,dN^l\big)\Big)\wedge * 
  \Big(d\tilde \xi_\lambda - \tfrac{1}2
  \I \big(\tilde K_{L^\lambda N^k}\,dN^k\big)\Big) \ .\nn
\eea
Since we intend to use these results in the effective action for Calabi-Yau
orientifolds, we make a further simplification. We demand that the kinetic potential
$\tilde K$ is only a function of $L^\lambda$ and the imaginary part of $N^k$, which we 
denote by $l^k=\I N^k$. This implies that all chiral fields $N^k$ admit a Peccei-Quinn 
shift symmetry acting on the real parts of $N^k$ as it is indeed the case for the
orientifold setups. Thus the effective Lagrangian \eqref{eff_act1} simplifies to 
\bea \label{linaction}
\cL &=& -\tfrac{1}{2}R*\mathbf{1} - 
   \tfrac{1}{4}\tilde K_{l^k l^l}\, dN^k \wedge * d \bar N^l
  + \tfrac{1}{4} \tilde K_{L^\kappa L^\lambda}\, 
  dL^\kappa \wedge * dL^\lambda \\ 
  && + 4 \tilde K^{L^\kappa L^\lambda}
  \Big(d\tilde \xi_\kappa + \tfrac{1}{4}\tilde K_{L^\kappa l^l}\, d\, \text{Re}N^l\Big)\wedge * 
  \Big(d\tilde \xi_\lambda + \tfrac{1}{4}\tilde K_{L^\lambda l^k}\, d\, \text{Re}N^k\Big) \ .\nn
\eea
This $N=1$ Lagrangian is written completely in terms of chiral multiplets and therefore 
can be derived from a K\"ahler potential when choosing appropriate complex coordinates
$N^k$ and $T_\lambda=(L^\lambda, \tilde \xi_\lambda)$.
As we will see in a moment, a direct calculation yields that this K\"ahler potential 
is the Legendre transform of $\tilde K$ with respect to the scalars $L^\kappa$. 
It takes the 
form 
\beq \label{LegKP}
 K(T,N) = \tilde K(L, N - \bar N)  - 2 (T_\kappa +\bar T_\kappa) L^\kappa
\eeq
where $L^\kappa(N,T)$ is a function of the complex fields $N^k,T_\lambda$. This 
dependence is implicitly given via the definition of the coordinates $T_\lambda$
\beq\label{defT}
T_\lambda = i\tilde \xi_\lambda + \tfrac{1}{4}\tilde K_{L^\lambda}\ . 
\eeq
However, in order to calculate the K\"ahler metric, one only needs to determine 
the derivatives of $L^\kappa(N,T)$ with respect to 
$N^k,T_\lambda$. They are obtained by differentiating \eqref{defT} and simply read 
\beq \label{derL}
  {\partial L^\kappa}/{\partial T_\lambda} = 2 \tilde K^{L^\kappa L^\lambda}\ , \qquad 
  {\partial L^\kappa}/{\partial N^l} = - \tfrac{1}{2i} \tilde K^{L^\kappa L^\lambda} \tilde K_{L^\lambda l^l}\ .
\eeq
Using these identities one easily calculates the first derivatives of the K\"ahler 
potential \eqref{LegKP} as 
\beq \label{Kder}
  K_{T_\alpha} = -2 L^\alpha\ , \qquad K_{N^A} = \tfrac{1}{2i} \tilde K_{l^A}\ .
\eeq
Applying the equations \eqref{derL} once more when differentiating \eqref{Kder} 
one finds the K\"ahler metric
\bea \label{Km1}
  K_{T_\alpha \bar T_\beta} &=& -4 \tilde K^{L^\alpha L^\beta}\ , \quad 
  K_{T_\alpha \bar N^A}\ =\ i \tilde K^{L^\alpha L^\beta} \tilde K_{L^\beta l^A}\ , \nn \\ 
  K_{N^A \bar N^B} &=& \tfrac{1}{4} \tilde K_{l^A l^B} - \tfrac{1}{4} 
                     \tilde K_{ l^A L^\alpha }\, \tilde K^{L^\alpha L^\beta}\, \tilde K_{L^\beta l^B}\ ,
\eea
with inverse
\bea \label{invKm1}
  K^{T_\alpha \bar T_\beta} &=& - \tfrac{1}{4} \tilde K_{L^\alpha L^\beta}
                    + \tfrac{1}{4} \tilde K_{ l^A L^\alpha }\, \tilde K^{l^A l^B} \, \tilde K_{L^\beta l^B}\ , 
                    \nn \\
  K^{T_\alpha \bar N^B} & = & -i \tilde K^{l^A l^B}\, \tilde K_{ l^A L^\alpha }\ , \quad 
  K^{N^A \bar N^B} \ = \ 4 \tilde K^{l^A l^B}\ .
\eea
Finally, one checks that $K(T,N)$ is indeed the K\"ahler potential for the chiral part of the 
Lagrangian \eqref{linaction}. This is done by inserting in the definition 
of $T_\kappa$ given in \eqref{defT} and the K\"ahler metric \eqref{Km1} into 
\beq
 \cL =  -\tfrac{1}{2}R*\mathbf{1} - K_{M^I \bar M^J}\ dM^I \wedge * d\bar M^J\ , 
\eeq
where $M^I=(N^k, T_\lambda)$.

\section{Gerneral reduction of the quaternionic space \label{Geom-of-M}}

In this appendix we present a more detailed analysis 
of the moduli space $\tilde \cM^{\rm Q}$, which is a K\"ahler
submanifold in the quaternionic space $\cM^{\rm Q}$. Our aim is to 
show that the K\"ahler potential \eqref{intKQ} with coordinates 
$T_\kappa,N^k$ introduced in \eqref{Oexp} indeed encode the correct 
low-energy dynamics of the theory obtained by Kaluza-Klein reduction.
Furthermore we show that $K^{\rm Q}$ always obeys a no-scale
type condition equivalent to \eqref{no-scale2}.
Most of the calculations will be based on the Legendre transform method 
applied to the real part of the coordinates $T_\kappa$. On the level of
superfields one can interpret this as dualization of these chiral 
multiplets into linear multiplets as discussed in appendix \ref{linm}.

Let us start by performing the reduction of the ten-dimensional 
theory by using the general basis $(\alpha_\Kh,\beta^\Kh)$ 
introduced in \eqref{decompO2}. It was chosen such that it splits on 
$H^3(Y)=H^{3}_+ \oplus H^3_-$ as 
\beq \label{basis1}
  (\alpha_k,\beta^\lambda) \in H^{3}_+(Y)\ , \qquad  (\alpha_\lambda,\beta^k) \in H^{3}_-(Y)\ ,
\eeq
where both eigenspaces are spanned by $h^{2,1}+1$ basis vectors.
As remarked above, we will only concentrate on the moduli space 
$\tilde \cM^{\rm Q}$, such that we can set $t^a=0$ and $A^\alpha=0$.
Due to \eqref{fieldtransf}, the ten-dimensional three-form $\hat C_3$ is expanded in 
elements of $H^{3}_+(Y)$ as 
\beq
  \CC_3 = \xi^k(x)\, \alpha_k - \tilde \xi_\lambda(x)\, \beta^\lambda \ ,
\eeq
where $\xi^k, \tilde \xi_\lambda$ are $h^{2,1}+1$ real space-time scalars in 
four-dimensions. Inserting this Ansatz into the ten-dimensional effective 
action one finds
\bea  \label{act2}
  S^{(4)}_{\tilde \cM^{\rm Q}} &=& \int -\, d D \wedge * dD  -\, G_{K L}(q)\, dq^K \wedge * dq^L 
         +\tfrac{1}{2} e^{2D}\, \text{Im}\, \cM_{ k  l}\, 
         d\xi^{ k} \wedge * d\xi^{l} \\
      &&  + \tfrac{1}{2} e^{2D}\, (\text{Im}\, \cM)^{-1\ \kappa \lambda}
      \big(d\tilde \xi_\kappa - \text{Re}\, \cM_{\kappa  l}\, 
      d\xi^{l} \big)
        \wedge * \big(d\tilde \xi_\lambda-\text{Re}\, \cM_{\lambda  k}\, d\xi^{ k} \big)\ , \nn 
\eea
where compared to \eqref{act1} only the terms involving $\xi^{k},\tilde \xi_\lambda$ have
changed. The metric $G_{K L}(q)$ was introduced in \eqref{def-G}
and is the induced metric on the space of real 
complex structure deformations $\cM^{\rm cs}_\bbR$ parameterized by $q^K$. 
It remains to comment on the kinetic and coupling terms of the 
scalars $\xi^k, \tilde \xi_\lambda$. In the quaternionic metric
\eqref{q-metr} of the $N=2$ theory they couple via the 
matrix $\cM_{\Kh \Lh}$ given in \eqref{defM}. Using the split of the symplectic basis 
$(\alpha_\Kh, \beta^\Kh)$ as given in \eqref{basis1} and the fact that
for $\alpha \in H^{3}_+, * \alpha \in H^{3}_-$ one concludes
\beq
  \text{Re} \cM_{\kappa \lambda}(q) = \text{Re} \cM_{k l}(q) = \text{Im} \cM_{\lambda k}(q) = 0\ , 
\eeq
whereas $\text{Re} \cM_{k \lambda}, \text{Im} \cM_{\kappa \lambda}, \text{Im} \cM_{k l}$ 
are generally non-zero on $\cM^{\text{cs}}_{\mathbb{R}}$. The explicit form of non-vanishing
components can be obtained by restricting \eqref{gauge-c} to $\cM^{\rm cs}_\bbR$ and
using the constraints \eqref{Z=0gen}.

In order to combine the scalars $e^D,q^K$ with $\xi^k, \tilde \xi_\lambda$ into 
complex variables, we have to redefine these fields and rewrite the first two 
terms in \eqref{act2}. Thus we define the $h^{2,1}+1$ real coordinates
\beq \label{lL-def}
   L^\lambda\ =\ - e^{2D}\, \I \big[C Z^\lambda(q) \big]\ ,\qquad 
   l^k \ =\  \R\big[C Z^k(q)\big]\ ,
\eeq
which is consistent with the orientifold constraint 
\eqref{Z=0gen}. The additional factor of $e^{2D}$ was included in order 
to match the dilaton factors later on.
Using \eqref{lL-def} one calculates the Jacobian matrix 
\beq
  \cS \ =\ 
  \left(
  \begin{array}{ccc} \partial L^\lambda/\partial e^{-D} &  \partial L^\lambda/\partial q^s &
   \partial L^\lambda/\partial q^\sigma\\
  \partial l^k/\partial e^{-D} &  \partial l^k/\partial q^s &
   \partial l^k/\partial q^\sigma\\
  \end{array} 
  \right)\ ,
\eeq 
where $q^K=(q^s,q^\sigma)$ are the $h^{(2,1)}$ real coordinates 
introduced in \eqref{embmap1}. One evaluates the derivatives by applying 
\eqref{constr2} such that 
\beq
  \cS \ =\ \left(
  \begin{array}{ccc}
   e^{3D} \I (CZ^\lambda) &
   - e^{2D} \I (C \chi^\lambda_{s} ) &
   - e^{2D} \R (C \chi^\lambda_{\sigma})\\
   e^{D}\R (CZ^k ) &
   \R (C\chi^k_{s}) &
   - \I \big(C \chi^k_{\sigma} \big) 
  \end{array} 
  \right)\ ,
\eeq
where $\chi^{\Lh}_K$ is defined in \eqref{def-chi}. It is now straight forward to 
rewrite \eqref{act2} by using the 
identities \eqref{spconst} of special geometry as
\bea \label{IIA1}
 S^{(4)}_{\tilde \cM^{\rm Q}} &=& \int  2 e^{-2D} \text{Im} \cM_{\kappa \lambda}\, dL^\kappa \wedge * dL^\lambda
                        + 2 e^{2D} \text{Im} \cM_{k l}\, dl^k \wedge * dl^l
                        + \tfrac{e^{2D}}{2}  \text{Im} \cM_{ k  l}\, 
                          d\xi^{ k} \wedge * d\xi^{l}  \nn\\   
      && + \tfrac{e^{2D}}{2} \, (\text{Im}\, \cM)^{-1\ \kappa \lambda}
      \big(d\tilde \xi_\kappa - \text{Re}\, \cM_{\kappa  k}\, 
      d\xi^{k} \big)
        \wedge * \big(d\tilde \xi_\lambda-\text{Re}\, \cM_{\lambda  k}\, d\xi^{ k} \big)\ . 
\eea
From \eqref{IIA1} one sees that the scalars $l^k$ and $\xi^k$ nicely combine 
into complex coordinates 
\beq
   N^k\ =\ \tfrac{1}{2}\xi^k +  i l^k\ =\ \tfrac{1}{2}\xi^k +  i \R(C Z^k)\ ,
\eeq
which corresponds to \eqref{def-NT}.
In contrast, one observes that 
the metric for the kinetic terms of the
scalars $\tilde \xi_\lambda$ is exactly the inverse of the one
appearing in the kinetic terms of the scalar fields $L^\lambda$. 
This hints to the fact that the Lagrangian 
\eqref{IIA1} is obtained by dualizing a set of linear multiplets 
$(L^\lambda, D^\lambda_2)$ into chiral multiplets 
$(L^\lambda,\tilde \xi_\lambda)$.
The effective action of several linear 
multiplets coupled to a set of chiral multiplets $N^k$ is 
given in equation \eqref{kinetic}. In analogy to the K\"ahler potential in 
the standard $N=1$ supergravity the couplings and kinetic terms of 
the linear and chiral multiplets are encoded by a single real function 
$\tilde K(L, N,\bar N)$ \cite{BGG}.  
Dualizing the massless two-forms $D^\lambda_{2}$ to scalar fields $\tilde \xi_\lambda$
as described in appendix \ref{linm} the resulting effective action in terms of 
$(L^\lambda,\tilde \xi_\lambda)$ and $N^k$  takes the form \eqref{linaction}.
This implies that \eqref{IIA1} is indeed obtained from \eqref{linaction},
when appropriately specifying the function $\tilde K$. 
To extract $\tilde K(L, N,\bar N)$ we compare the action \eqref{IIA1} with 
\eqref{linaction} and read off the metric
\beq \label{lLmetric}
   \tilde K_{L^\kappa  L^\lambda}\ =\ 8\, e^{-2D} \IM_{\kappa \lambda}\ , \quad 
   \tilde K_{l^k l^l}\ =\ -8\, e^{2D} \IM_{k l} \ , \quad 
   \tilde K_{L^\kappa  l^l} \ =\  - 8\, \RM_{\kappa l}\ ,
\eeq
where we have used that the metric is independent of $\xi^k,\tilde \xi_\lambda$.
This metric can be obtained from a kinetic potential of the form
\beq \label{kinpo1}
  \tilde K(L,l)\ =\ - \ln\big[ e^{-4D} \big]+ 8e^{2D}\I \big[\rho^*\cF(CZ^k)\big]\ ,
\eeq
where $\cF$ is the prepotential of the special K\"ahler manifold $\cM^{\rm cs}$
restricted to the real subspace $\cM^{\rm cs}_\bbR$. The map $\rho$ was given 
in \eqref{embmap1} and enforces the constraints \eqref{Z=0gen}. To show that $\tilde K$
indeed yields the correct metric \eqref{lLmetric} one differentiates \eqref{kinpo1}
with respect to $e^{-D},q^K$ and uses the inverse of $\cS$. Applying equations
\eqref{ML-hf} one finds its first derivatives  
\beq \label{first-der}
  \tilde K_{ L^\lambda} \ =\ - 8\, \R\big[C \cF_\lambda(q) \big] \qquad
   \tilde K_{l^\kh} \ = \ 8\, e^{2D}\, \I\big[C \cF_{k}(q) \big]\ .
\eeq
Repeating the procedure and differentiating \eqref{first-der}
with respect to $e^{-D},q^K$ and using $\cS^{-1}$ one can apply
\eqref{def-M} to show \eqref{lLmetric}. 

As explained in appendix \ref{linm} the actual K\"ahler potential of 
$\tilde \cM^{\rm Q}$ is the Legendre transform of $\tilde K$ with 
respect to the variables $L^\lambda$. There we also found the explicit 
definition of the complex coordinates $T_\lambda$ combining $(L^\lambda,\tilde \xi_\lambda)$.  
Thus the K\"ahler potential $K^{\rm Q}(T,N)$ is obtained from $\tilde K(L,N)$
by setting
\beq \label{K-Legendre}
  K^{\rm Q}(T,N)\ =\tilde K(L,N) - 2(T_\lambda + \bar T_\lambda) L^\lambda\ ,
\eeq
where $L^\lambda(T+\bar T,N,\bar N)$ is now a function of the chiral multiplets $T_\lambda$ and
$N^k$. This dependence is implicitly defined via the equation 
\beq
  T_\lambda + \bar T_\lambda = \tfrac{1}{2} \tilde K_{L^\lambda}\ . 
\eeq
Using \eqref{first-der} and fixing the normalization of the 
imaginary part of $T_\lambda$ by comparing \eqref{IIA1} with \eqref{linaction}
one finds
\beq
  T_\lambda = i \tilde \xi_\lambda + \tfrac{1}{4}\tilde K_{L^\lambda} 
           = i \tilde \xi_\lambda - 2\, \R\big(C F_\lambda\big) \ ,
\eeq
which coincides with \eqref{def-NT} already quoted in section \ref{eff_supform}.
To give an explicit expression for  $K^{\rm Q}$ we plug  
equation \eqref{kinpo1} into \eqref{K-Legendre}. Inserting the $N=2$ identity 
$\cF=\frac12 Z^\Kh \cF_\Kh$, the constraint equations \eqref{Z=0gen} 
and \eqref{lL-def},\eqref{first-der} we rewrite 
\beq  \label{K_lL}
    K^{\rm Q}= - \ln\, e^{-4D} + \tfrac{1}{2} (l^k \tilde K_{l^k} - L^\lambda \tilde K_{ L^\lambda})\ .
\eeq
It is possible to evaluate the terms appearing in the parentheses. In order to do that 
we combine the equations \eqref{lL-def} and \eqref{first-der} to the simple form 
\bea \label{lL}
  \R\big( C \Omega \big)\ =\ l^k \alpha_k + \tfrac{1}{8} \tilde K_{L^\lambda} \beta^\lambda\ ,\quad
   e^{2D} \I\big( C \Omega \big)\ =\ -L^\lambda \alpha_\lambda - \tfrac{1}{8} \tilde K_{l^k} \beta^k\ . 
\eea
We now use equation \eqref{csmetric} and the definition \eqref{def-C} of $C$
to calculate
\beq \label{skconstr}
 2 \int_Y \R( C\Omega) \wedge \I(C\Omega) = i \int_Y C\Omega \wedge \overline{C\Omega} = e^{-2D}\ .
\eeq
Inserting the equations \eqref{lL} into \eqref{skconstr} we find 
\bea \label{lL=4} 
  L^\lambda \tilde K_{L^\lambda} - l^k \tilde K_{l^k} = 4\ .
\eea
Inserting this constraint into \eqref{K_lL} we have shown that the K\"ahler potential
has indeed the form \eqref{intKQ}.\footnote{By using the equation \eqref{skconstr} and $*\Omega=-i\Omega$ 
it is straight forward to show $e^{-2D}=2\int \R(C\Omega)\wedge * \R(C\Omega)$}
Moreover, \eqref{lL=4} directly translates into a no-scale type condition for $K^{\rm Q}$
\bea \label{no-scale4}
  K_{w^\Kh} K^{w^\Kh \bar w^\Lh} K_{\bar w^\Lh} = 4\ ,
\eea
where $w^\Kh=(T_\kappa, N^k)$.
In order to see this, one inserts the inverse K\"ahler metric \eqref{invKm1},
the K\"ahler derivatives \eqref{Kder} and the derivatives of \eqref{lL=4} back into
\eqref{lL=4}. In other words, we were able to translate one of 
the special K\"ahler conditions present in the underlying 
$N=2$ theory into a constraint on the geometry of 
$\tilde \cM^{\rm Q}$. 

Let us end our discussion of $\tilde \cM^{\rm Q}$, by giving two specific 
examples for $\tilde K$ satisfying the constraint \eqref{lL=4}, namely
\bea
  \tilde K_1&=& \ln \Big[a_1 \frac{\kappa_{\kappa \lambda \mu} L^\kappa L^\lambda L^\mu }{l^0} \Big] 
             + b_1 \frac{\kappa_{\kappa kl } L^\kappa l^k l^l}{l^0} \ , \nn \\
  \tilde K_2&=& -\ln \Big[a_2 \frac{\kappa_{klm} l^k l^l l^m }{L^0} \Big] 
             +  b_2 \frac{\kappa_{\kappa \lambda l } L^\kappa L^\lambda l^l}{L^0}\ ,
\eea
where $a_{1,2},b_{1,2}$ are some constants. In \cite{GL} it was shown, that $\tilde K_1$ 
is the correct potential describing the dynamics of IIB orientifolds
with $O3/O7$ planes. On the other hand, $\tilde K_2$ is the correct potential for 
IIB orientifolds with $O5/O9$ planes. $\tilde K_{1,2}$ have this 
simple form since instanton corrections are not taken into account. 


\section{The Geometry of the moduli space of CY orientifolds \label{Geom_modspace}}

In this section we give an alternative formulation of 
the geometric structures of the moduli space $\tilde \cM^{\rm Q}$ 
which is closely related the moduli space of 
supersymmetric Lagrangian submanifolds in a Calabi-Yau 
threefold \cite{Hitchin2}.\footnote{This 
analysis can equivalently be applied to the moduli space of 
$G_2$ compactifications of 
M-theory.}  In this set-up also
the no-scale conditions \eqref{no-scale2}, 
\eqref{lL=4} are interpreted geometrically.

In section~\ref{eff_supform} we started from an $N=2$ quaternionic
manifold $\cM^{\rm Q}$ and determined the submanifold
$\tilde\cM^{\rm Q}$ by imposing the orientifold projection.
$N=1$ supersymmetry ensured that this submanifold is K\"ahler.
$\cM^{\rm Q}$ has a second but different K\"ahler submanifold
$\cM^{\rm cs}$ which intersects with $\tilde\cM^{\rm Q}$
 on the real manifold $\cM^{\rm cs}_\bbR$.
The c-map is in some sense the reverse operation where 
$\cM^{\rm Q}$ is constructed starting from $\cM^{\rm cs}$
and shown to be quaternionic \cite{CFGi,FS}.
In this appendix we analogously construct the K\"ahler manifold 
$\tilde\cM^{\rm Q}$
starting from 
$\cM^{\rm cs}_\bbR$.

In fact the proper starting point is not $\cM^{\rm cs}_\bbR$ but rather
$\cM_\bbR=\cM_\bbR^{\rm cs} \times \bbR$ which is the local product of the
moduli space 
of real complex structure deformations of a Calabi-Yau orientifold 
times the real dilaton direction.\footnote{The $N=2$ analog of 
$\cM_\bbR$ is the extended moduli space 
$\hat\cM^{\rm cs} = \cM^{\rm cs} \times \bbC$ where $\bbC$ 
is the complex line normalizing $\Omega$. The corresponding modulus
can be identified with the complex dilaton \cite{Witten2}.
The orientifold projection fixes the phase of the complex dilaton
(it projects out the four-dimensional $B_2$) to be $\theta$ and thus reduces
$\bbC$ to   $\bbR$. }
Its local geometry is encoded in the variations of the real and imaginary part of  the normalized holomorphic three-form $C\Omega$.
This form naturally defines an embedding 
\beq
  E: \cM_\bbR \rightarrow V \times V^*
= \ H^3_+(\mathbb{R}) \times H^3_-(\mathbb{R})\ .
\eeq
where $V =H^3_+(\mathbb{R})$ and we used the intersection 
form $\big<\alpha,\beta \big>=\int \alpha \wedge \beta $ on $H^{3}(Y)$ 
to identify $V^* \cong H^3_-(\mathbb{R})$. 
 $V \times V^*$ naturally admits a
symplectic form $\cW$ and an indefinite metric $\cG$ defined as
\bea
  \cW((\alpha_+,\alpha_-),(\beta_+,\beta_-)) = \big<\alpha_+, \beta_-\big> - \big<\beta_+ ,\alpha_-\big>\ , \nn\\
  \cG((\alpha_+,\alpha_-),(\beta_+,\beta_-)) = \big<\alpha_+, \beta_-\big> + \big<\beta_+ ,\alpha_-\big>\ ,
\eea
where $\alpha_\pm,\beta_\pm \in H^3_\pm(\mathbb{R})$.

Now we construct $E$ in such a way that 
$\cM_{\bbR}$ is a Lagrangian submanifold of $V \times V^*$ with respect to 
$\cW$ and its metric is induced from $\cG$, i.e.\ 
\beq\label{Lagr}
 E^*(\cW)=0 \ , \qquad E^*(\cG)=g
\eeq
where
\beq \label{metrQ}
  \tfrac{1}{2} g = dD \otimes dD +  G_{K L} d q^K \otimes d q^L\ .
\eeq 
 is the metric on $\cM_\bbR$ as determined in \eqref{act1}.
As we are going to show momentarily $E$ is given by
\bea  \label{embmap}
   E(q^\Kh) = \big(2\R(C\Omega)\, ,\ -2e^{2D}\I(C\Omega)\big)\ ,
\eea 
where $q^\Kh=(e^{-D},q^K)$ and 
$\Omega$ is evaluated at  $q^K \in \cM^{\rm cs}_\bbR$.
Additionally $E$ satisfies
\beq \label{no-scale3}
  \cG(E(q^\Kh),E(q^\Kh)) = 4\ ,
\eeq
for all $q^K $. This implies that the image of all points in 
$\cM_\bbR$ have the same distance from the origin. Later on we will show that
this translates into the no-scale condition \eqref{no-scale4}.

Before we do so 
let us first show that the $E$ given in \eqref{embmap} indeed satisfies
\eqref{Lagr} and \eqref{no-scale3}. 
The explicit calculation is straight forward and essentially included in the 
reduction presented in appendix \ref{Geom-of-M}.\footnote{Formally one has to  
first evaluate $E_* (\partial_{Q^\Kh})$ and expresses the result in terms of the $(3,0)$-form $\Omega$ 
and the $(2,1)$-forms $\chi_K$. One than uses that by definition of the pullback 
$E^* \omega(\partial_{q^\Kh},\cdot ) = \omega(E_* (\partial_{q^\Kh}),\cdot)$ for a form $\omega$ on $V \times V^*$. 
Applied to $\cG$ and $\cW$ one finds that the truncation of the special K\"ahler 
\eqref{csmetric} and \eqref{chi_barchi} indeed imply \eqref{Lagr}. This calculation does not
make use of any specific basis of $H^3_\pm$.} To see this we 
express the map $E$ defined in \eqref{embmap} and the conditions in terms of the basis 
$(\alpha_\Kh,\beta^\Kh)$ introduced in \eqref{basis1}. We use eq. \eqref{lL} and expand 
\beq \label{Eincoords}
  E(q^\Kh)\ =\ \big(2l^k \alpha_k + \tfrac{1}{4} \tilde K_{L^\kappa} \beta^\kappa, 
                2L^\kappa \alpha_\kappa + \tfrac{1}{4} \tilde K_{l^k} \beta^k\big)\ ,
\eeq
where $l^k,L^\kappa$ and $\tilde K_{L^\kappa}, \tilde K_{l^k}$ are functions of $q^\Kh$ as given in 
\eqref{lL-def} and \eqref{first-der}.
We define coordinates $u^\Kh=(2l^k, \tfrac{1}{4}\tilde K_{L^\kappa})$ on $V$ and 
coordinates $v_\Kh=(\tfrac{1}{4}\tilde K_{l^k},-2L^\kappa)$ on $V^*$. 
In these coordinates the first two conditions in \eqref{Lagr} simply read
\beq \label{Lagrc}
    E^*(du^\Kh \wedge dv_\Kh)=0\ , \qquad E^*(du^\Kh \otimes dv_\Kh) = g\ .
\eeq
{}From appendix \ref{Geom-of-M} we further know that 
$\tilde K_{L^\kappa}, \tilde K_{l^k}$ are derivatives 
 of a kinetic potential $\tilde K$ and thus we can evaluate $du^\Kh$ 
and $dv_\Kh$ in terms of $l^k,L^\kappa$.
Inserting the result into \eqref{Lagrc} 
the second equation can be rewritten as
\beq
  \tfrac{1}{2} g\ =\ \tfrac{1}{4} 
        \tilde K_{l^k l^l}\, dl^k \otimes dl^l - 
        \tfrac{1}{4} 
        \tilde K_{L^\kappa L^\lambda}\, dL^\kappa \otimes dL^\lambda\ ,
\eeq 
while the first equation is trivially fulfilled due to the symmetry of $\tilde K_{l^k l^l}$
and $\tilde K_{L^\kappa L^\lambda}$. This metric is exactly the one appearing in the action 
\eqref{IIA1} when using \eqref{lLmetric}. Expressing $g$ in coordinates 
$e^{D},q^K$ leads to \eqref{metrQ}, as we have already checked by going from 
\eqref{act2} to \eqref{IIA1} above.
Furthermore, inserting \eqref{Eincoords} into \eqref{no-scale3} 
it exactly translates 
into the no-scale condition \eqref{lL=4}, which was shown in appendix
\ref{Geom-of-M} to be equivalent to \eqref{no-scale2}.

We have just shown that $\cM_\bbR$ is a 
Lagrangian submanifold of  $V \times V^*$.
Identifying $T^*V \cong V \times V^*$ we conclude that  $\cM_\bbR$ 
can be obtained as the graph $(\alpha(u),u)$ 
of a closed one-form $\alpha$. This implies that we can locally find 
a generating function $K': V \rightarrow \mathbb{R}$ such that $\alpha = dK'$. 
In local coordinates $(v_\Kh,u^\Kh)$ this amounts to
\beq \label{v=K/u}
  v_\Kh = \frac{\partial K'}{\partial u^{\Kh}}
\eeq 
such that 
\beq
  - L^\kappa(u) = 2\, \frac{\partial K'(u)}{\partial \tilde K_{L^\kappa}}\ , \quad 
  \tilde K_{l^k}(u) = 2\, \frac{\partial K'(u)}{\partial l^k}\ . 
\eeq
These equations are satisfied if we define $K'$ in terms of $\tilde K$ as
\bea \label{Legendre}
  2 K'\ =\ \tilde K(L(u),l) - \tilde K_{L^\kappa}(u)\, L^\kappa(u)\ , 
\eea  
which is nothing but the Legendre transform of $\tilde K$ with respect to $L^\kappa$.
Later on we show that the function ${2}K'$  is identified with the K\"ahler potential 
$K$ given in \eqref{intKQ}. 

In order to do that, we now extend our discussion to the 
full moduli space $\tilde \cM^{\rm Q}$ including the scalars 
$\zeta^\Kh=(\xi^k,\tilde \xi_\kappa)$ parameterizing the 
three-form $\hat C_3$ in $H^{3}_+(\bbR)$. Locally one has
\beq
   \tilde \cM^{\rm Q} = \cM_\bbR \times H^{3}_+(\bbR)\ .
\eeq
The tangent space at a point $p$ in $ \tilde \cM^{\rm Q}$ can be identified as
\beq
  T_p \tilde \cM^{\rm Q} \cong H^3_+(\bbR)\oplus H^3_+(\bbR) \cong H^3_+(\bbR) \otimes \bbC\ ,
\eeq
where the first isomorphism is induced by the embedding $E$ given in \eqref{embmap}.
This is a complex vector space and thus $\tilde \cM^{\rm Q}$ admits an 
almost complex structure $I$. In components it is given by
\beq \label{def-I}
  I(\partial_{q^\Kh}) = (\partial u^\Lh/\partial q^\Kh)\, \partial_{\zeta^\Lh}\ ,\qquad
  I((\partial u^\Lh/\partial q^\Kh)\, \partial_{\zeta^\Lh})=-\partial_{q^\Kh}\ ,
\eeq
where we have used that $I$ is induced by the embedding map $E$. One can show that
the almost complex structure $I$ is integrable, since  
\beq
  dw^\Kh = du^\Kh + i d \zeta^\Kh = (\partial u^\Lh/\partial q^\Kh) dq^\Kh + i d \zeta^\Kh\ ,
\eeq
are a basis of $(1,0)$ forms and $w^\Kh=u^\Kh+i\zeta^\Kh$ are complex coordinates on $\tilde \cM^{\rm Q}$. 
Using the definition of $u^\Kh$ one infers that as expected $w^\Kh = (N^k,T_\kappa)$. 
Moreover, one naturally extends the metric $g$ on $T \cM_\bbR$ to a hermitian metric 
on $T\tilde \cM^{\rm Q}$. The corresponding two-form is then given by
\beq
  \tilde \omega(\partial_{\zeta^\Lh}, \partial_{q^\Kh}) = g(I\partial_{\zeta^\Lh},\partial_{q^\Kh})\ ,
  \qquad \tilde \omega(\partial_{\zeta^\Kh},\partial_{\zeta^\Lh})= \tilde \omega(\partial_{q^\Kh},\partial_{q^\Lh})=0\ .
\eeq
Using the definition \eqref{def-I} of the almost complex structure and 
equation \eqref{Lagr}, one concludes that $\tilde \omega$ is given by
\beq \label{tildeo}
   \tilde \omega= dv_\Kh \wedge d\zeta^\Kh 
                = 2i \frac{\partial^2 K'}{\partial w^\Kh \partial \bar w^\Lh} dw^\Kh \wedge d\bar w^\Lh   \ ,
\eeq
where for the second equality we applied \eqref{v=K/u} and expressed the 
result in coordinates $w^\Kh=u^\Kh + i \zeta^\Kh$. Note that $K'$ is a function 
of $u^\Kh$ only, such that derivatives with respect to $w^\Kh$ translate to the ones
with respect to $u^\Kh$. Equation \eqref{tildeo} implies that $K^{\rm Q}=2K'$ is indeed
the correct K\"ahler potential for the moduli space $\tilde \cM^{\rm Q}$.


\end{document}